\documentclass[11pt]{article}

\usepackage{amssymb}
\usepackage{amsthm}
\usepackage{amsmath}

\usepackage[mathscr]{eucal}

\usepackage{graphicx}
\usepackage{psfrag}
\usepackage{subfigure}

\newcommand{\proofend}{\hspace*{\fill}\rule{0.2cm}{0.2cm}}

\newcommand{\textfrac}[2]{{\textstyle{\frac{#1}{#2}}}}
\newcommand{\sfrac}[2]{\textfrac{#1}{#2}}

\newcommand{\weg}{\:\,}

\newcommand{\spur}{\mathrm{tr}\hspace{0.15ex}}

\newcommand{\cD}{\mathcal{D}}

\newcommand{\cH}{\mathcal{H}}
\newcommand{\cK}{\mathcal{K}}
\newcommand{\cS}{\mathcal{S}}
\newcommand{\cZ}{\mathcal{Z}}
\newcommand{\cX}{\mathcal{X}}

\theoremstyle{plain}
\newtheorem{theorem}{Theorem}[section]

\newtheorem{lemma}[theorem]{Lemma}

\theoremstyle{remark}

\newtheorem*{remark}{Remark}

\begin{document}

\title{\huge \sc Dynamics of the spatially homogeneous Bianchi type
I Einstein-Vlasov equations}

\author{ \\
{\Large\sc J.\ Mark Heinzle}\thanks{Electronic address:
{\tt Mark.Heinzle@aei.mpg.de}} \\[1ex]
Max Planck Institute for Gravitational Physics, \\
Albert Einstein Institute, \\
Am M\"uhlenberg 1, D-14476 Golm, Germany
\and \\
{\Large\sc Claes Uggla}\thanks{Electronic address:
{\tt claes.uggla@kau.se}} \\[1ex]
Department of Physics, \\
University of Karlstad, S-651 88 Karlstad, Sweden \\[2ex] }

\date{}
\maketitle
\begin{abstract}
We investigate the dynamics of spatially homogeneous solutions of
the Ein\-stein-Vlasov equations with Bianchi type I symmetry by
using dynamical systems methods. All models are forever expanding
and isotropize toward the future; toward the past there exists a
singularity. We identify and describe all possible past asymptotic
states; in particular, on the past attractor set we establish the
existence of a heteroclinic network, which is a new type of feature
in general relativity.
This illustrates among other things that
Vlasov matter can lead to quite different
dynamics of cosmological models as compared to perfect fluids.

\end{abstract}
\centerline{\bigskip\noindent PACS numbers: 04.20.Dw, 98.80.Bp}

\vfill
\newpage

\section{Introduction}
\label{introduction}

In general relativity and cosmology, our knowledge about spatially
homogeneous cosmological models has increased substantially over the
years, and we are able to say that, for a large number of models,
the qualitative behaviour of solutions is now well understood;
see~\cite{waiell97} for an overview. The majority of results,
however, concerns solutions of the Einstein equations coupled to a
perfect fluid (usually with a linear equation of state). It is thus
important to note that these results are in general not robust,
i.e., not structurally stable, under a change of the matter model;
significant changes of the qualitative behaviour of solutions occur,
for instance, for collisionless matter.

Several fundamental results on spatially homogeneous diagonal models of
Bianchi type~I with collisionless matter have been obtained in~\cite{ren96}.
Diagonal locally rotationally symmetric (LRS) models have been investigated
successfully by using dynamical systems methods, see~\cite{rentod99} for the
case of massless particles, and~\cite{renugg00,ren02} for the massive case.
In particular, solutions have been found whose qualitative behaviour is
different from that of any perfect fluid model of the same Bianchi type.

The purpose of this article is to re-investigate the diagonal (non-LRS)
Bianchi type~I models with collisionless matter. Our analysis is based on
dynamical systems techniques, which enables us to obtain a more detailed
picture of the global dynamics than the one previously given.
In particular, we show that the dynamical behaviour toward
the past singularity of the collisionless matter model
differs considerably from the Bianchi type~I perfect fluid model.

The outline of the paper is as follows. In Section~\ref{einsteinvlasov}
we reformulate Einstein's field equations for the diagonal Bianchi
type~I case with collisionless matter as a reduced 
dimensionless dynamical system on a compact state space. In
Section~\ref{invfixed} we give the fixed points of the system and list
and discuss a hierarchy of invariant subsets of the state space,
which is associated with a hierarchy of monotone functions. In
Section~\ref{locglo} we first present the results of the local
dynamical systems analysis; subsequently, we focus on
the analysis of the global dynamics: we establish two theorems that
formulate the future and past asymptotic behaviour of solutions,
respectively. As regards the future asymptotics we show that all
models isotropize asymptotically toward the future. The past
asymptotic behaviour is more complicated since there exists several
types of past asymptotic behaviour; in particular we establish that
the past attractive set resides on a set that contains a so-called
heteroclinic network. The proof of the theorems is based on methods
from global dynamical systems analysis; in particular, we exploit
the hierarchy of monotone functions in conjunction with the
monotonicity principle. Finally, in Section~\ref{conc} we conclude with
some remarks about our results and their implications.
Appendix~\ref{dynsys} provides a brief introduction to relevant
background material from the theory of dynamical systems; in
particular we cite the monotonicity principle. The proofs of some of
the statements in the main text are given in Appendix~\ref{FRWLRS}
and~\ref{futureproof}. In Appendix~\ref{Si0space} we discuss the
physical interpretation of one of the most important boundaries of
our state space formulation.

\section{The reflection-symmetric Bianchi type~I Einstein-Vlasov system}
\label{einsteinvlasov}

In a spacetime with Bianchi type~I symmetry the spacetime metric can
be written as
\begin{equation}
d s^2 = -d t^2 + g_{i j}(t) d x^i d x^j\:,\quad i,j=1,2,3\:,
\end{equation}
where $g_{i j}$ is the induced Riemannian metric on the spatially
homogeneous surfaces $t=\mathrm{const}$. Since the metric is
constant on $t=\mathrm{const}$, it follows that the Ricci tensor of
$g_{i j}$ vanishes. Einstein's equations, in units $G=1=c$,
decompose into the evolution equations,
\begin{subequations}\label{einsteinvlasovsystem}
\begin{equation}\label{evolution}
\partial_t g_{i j} = -2 k_{i j} \:,\quad
\partial_t k^i_{\weg j} = \spur k\: k^i_{\weg j} - 8 \pi T^i_{\weg j} +
4 \pi \delta^i_{\weg j} ( T^k_{\weg k} -\rho) - \Lambda \delta^i_{\weg j}\:,
\end{equation}
and the Hamiltonian and momentum constraint
\begin{equation}\label{constraints} (\spur k)^2
- k^i_{\weg j} k^j_{\weg i} - 16 \pi \rho -2 \Lambda = 0\:,\qquad
j_k = 0\:.
\end{equation}
Here, $k_{i j}$ denotes the second fundamental form of the surfaces $t=\mathrm{const}$.
The matter variables are defined as components of the energy-momentum
tensor $T_{\mu\nu}$ ($\mu=0,1,2,3$), according to $\rho = T_{00}$,
$j_k= T_{0k}$; $T_{i j}$ denotes the spatial components.
The cosmological constant $\Lambda$ is set to zero in the following;
the treatment of the case $\Lambda>0$ is straightforward once the case $\Lambda=0$ has been solved,
cf.~the remarks in the conclusions.

In this paper we consider collisionless matter (Vlasov matter),
i.e., an ensemble of freely moving particles described by a
non-negative distribution function $f$ defined on the mass shell bundle
$PM\subseteq TM$ of the spacetime; for simplicity we consider
particles with equal mass $m$.
The spacetime coordinates $(t,x^i)$
and the spatial components $v^i$ of the four-momentum $v^\mu$
(measured w.r.t.\ $\partial/\partial x^\mu$) provide local
coordinates on $PM$ so that $f = f(t,x^i,v^j)$. Compatibility with
Bianchi type~I symmetry forces the distribution function $f$ to
be homogeneous, i.e., $f = f(t, v^j)$. The evolution equation for
$f$ is the Vlasov equation (the Liouville equation)
\begin{equation}\label{Vlasovequation}
\partial_t f + \frac{v^j}{v^0} \partial_{x^j} f -
\frac{1}{v^0} \Gamma^j_{\mu\nu} v^\mu v^\nu \partial_{v^j} f  =
\partial_t f + 2 k^j_{\weg l} v^l \partial_{v^j} f = 0 \:.
\end{equation}
The energy-momentum tensor associated with the distribution
function $f$ is given by
\[
T^{\mu\nu} = \int f v^\mu v^\nu \mathrm{vol}_{PM}\:,
\]
where $\mathrm{vol}_{PM} =  (\det g)^{1/2} v_0^{-1} d v^1 d v^2 d
v^3$ is the induced volume form on the mass shell; $v_0$ is
understood as a function of the spatial components, i.e., $v_0^2 =
m^2 + g_{i j} v^i v^j$. The components $\rho$, $j_k$, and $T_{i j}$,
which enter in~\eqref{evolution} and~\eqref{constraints} can thus be
written as
\begin{align}
\label{matterrho}
\rho & = \int f \left(m^2 + g^{i j} v_i v_j\right)^{1/2}
(\det g)^{-1/2} d v_1 d v_2 d v_3\:, \\
\label{matterj}
j_k & = \int  f v_k (\det g)^{-1/2} d v_1 d v_2 d v_3 \:,\\
\label{matterT}
T_{i j} & = \int f v_i v_j
\left(m^2 + g^{k l} v_k v_l\right)^{-1/2}
(\det g)^{-1/2} d v_1 d v_2 d v_3 \:.
\end{align}
\end{subequations}
The Einstein-Vlasov system~\eqref{einsteinvlasovsystem} is usually
considered for particles of mass $m>0$, however, the system also
describes massless particles if we set $m=0$. (For a detailed
introduction to the Einstein-Vlasov system we refer to~\cite{and05}
and~\cite{ren04}.)

The general spatially homogeneous solution
of the Vlasov equation~(\ref{Vlasovequation}) in
Bianchi type~I is
\begin{equation}\label{fisf0}
f(t,v^i) = f_0(v_i)\:,
\end{equation}
where the $v_i$ are the covariant components of the momenta and
$f_0$ is an arbitrary non-negative function, see~\cite{maamah90}.
(By inserting~(\ref{fisf0})
into~(\ref{Vlasovequation}) and using that $v_i = g_{i j}(t) v^j$
it is easy to check that $f_0(v_i)$ is a solution.)
The momentum constraint in~(\ref{constraints}) then reads
\begin{equation}\label{momcons}
\int f_0(v_i) v_k  d v_1 d v_2 d v_3 = 0\:.
\end{equation}
Henceforth, for simplicity, $f_0$ is assumed to be compactly supported.

There exists a subclass of Bianchi type~I Einstein-Vlasov models
that is naturally associated with the constraint~\eqref{momcons}:
the class of ``reflection-symmetric'' (or ``diagonal'') models.
The following symmetry conditions are imposed on the initial data:
\begin{subequations}\label{reflsymm}
\begin{equation}\label{reflsymmf0}
f_0(v_1, v_2,v_3) = f_0(-v_1,-v_2,v_3) = f_0(-v_1,v_2,-v_3) = f_0(v_1,-v_2,-v_3)\:,
\end{equation}
\begin{equation}\label{reflsymmgk}
g_{i j}(t_0)\:, k_{i j}(t_0) \quad\text{diagonal}\:.
\end{equation}
\end{subequations}
These conditions ensure that $T_{i j}(t_0)$ is diagonal, hence $g_{i
j}$, $k_{i j}$, and $T_{i j}$ are diagonal for all times by~(\ref{einsteinvlasovsystem}). 
In the present paper, we will be
concerned with this class of reflection-symmetric models.

The Einstein-Vlasov system~\eqref{einsteinvlasovsystem} thus reduces
to a system for six unknowns, the diagonal components of the metric
$g_{i i}(t)$ and the second fundamental form $k^i_{\weg i}(t)$ (no
summation). The equations are~\eqref{evolution} and the Hamiltonian
constraint in~\eqref{constraints}. The initial data consists of
$g_{i i}(t_0)$, $k^i_{\weg i}(t_0)$; in addition we prescribe a
distribution function $f_0(v_i)$ that provides the source terms in
the equations via~\eqref{matterrho} and~\eqref{matterT}.

In the following we reformulate the Einstein-Vlasov system as a
dimensionless system on a compact state space.
To that end we introduce new variables and modified matter quantities. Let
\begin{equation}\label{hx}
H := -\frac{\spur k}{3}\:, \quad\qquad x := \sum_i g^{i i}\:,
\end{equation}
and define the dimensionless variables
\begin{subequations}
\begin{align}
\label{defdimless}
& s_i := \frac{g^{i i}}{x}\; , & & \Sigma_i := -\frac{k^i_{\weg i}}{H} - 1\; ,
& z & := \frac{m^2}{m^2 + x}\:, \\[1ex]
\text{where} \qquad & \!\sum_i s_i =1\; , & & \sum_i\Sigma_i = 0\:.
\end{align}
\end{subequations}
The transformation from the variables $(g_{ii}, k^i_{\weg i})$ to
$(s_i,\Sigma_i, x, H)$, where $(s_i,\Sigma_i)$ are subject to the
above constraints, is one-to-one. (Note that $x$ can be obtained
from $z$ when $m>0$.) By distinguishing one direction ($1$, $2$, or
$3$), one can decompose the $s_i$ and simultaneously introduce a
trace-free adaption of the shear to new $\Sigma_\pm$ variables as is
done in, e.g.,~\cite{waiell97}; however, since Bianchi type~I does
not have a preferred direction we will refrain from doing so here.

Next we replace the matter quantities $\rho$, $T^i_{\weg i}$ (no
summation) by the dimensionless quantities
\begin{equation}
\Omega :=\frac{8\pi\rho}{3 H^2}\,,\qquad w_i := \frac{T^i_{\weg
i}}{\rho}\,,\qquad w := \frac{1}{3} \sum_i w_i = \frac{1}{3}
\frac{\sum_i T^i_{\weg i}}{\rho}\,.
\end{equation}
Expressed in the new variables, $w_i$ can be written as
\begin{equation}\label{omegai}
w_i = \frac{(1-z) s_i {\displaystyle\int} f_0\, v_i^2
\left[z+(1-z) \sum_k s_k \, v_k^2\right]^{-1/2} d v_1 d v_2 d v_3}%
{{\displaystyle\int} f_0 \left[z+(1-z) \sum_k s_k \,
v_k^2\right]^{1/2} d v_1 d v_2 d v_3}\:.
\end{equation}

Finally, let us introduce a new dimensionless time variable $\tau$ defined by
\begin{equation}
\partial_\tau = H^{-1}\partial_t\:;
\end{equation}
henceforth a prime denotes differentiation w.r.t.\ $\tau$.

We now rewrite the Einstein-Vlasov equations as a set of dimensional
equations that decouple for dimensional reasons and a reduced system
of dimensionless coupled equations on a compact state space. The
decoupled dimensional equations are
\begin{subequations}
\begin{align}
\label{Heq}
H^\prime &= -3 H \left[1 -\frac{\Omega}{2} (1-w)\right]\\
\label{xeq}
x^\prime &= -2 x \left(1 + \sum_k \Sigma_k s_k \right)\:.
\end{align}
\end{subequations}
The reduced dimensionless system consists of the Hamiltonian
constraint, cf.~\eqref{constraints},
\begin{equation}\label{omega}
1- \Sigma^2-\Omega = 0\:, \qquad\text{where}\quad \Sigma^2 :=
\sfrac{1}{6} \sum_k \Sigma_k^2\:,
\end{equation}
which we use to solve for $\Omega$,
and a coupled system of evolution equations
\begin{subequations}\label{eq}
\begin{align}
\label{Sigeq}
\Sigma_i^\prime & = -3 \Omega \left[ \frac{1}{2} (1-w) \Sigma_i -(w_i - w)\right]\\
\label{seq}
s_i^\prime & = -2 s_i \left[\Sigma_i - \sum_k \Sigma_k s_k \right] \\
\label{zeq}
z^\prime & = 2 z\,(1 - z)\left(\,1 + \sum_k s_k \, \Sigma_k\,\right)\:.
\end{align}
\end{subequations}
In the massive case $m>0$ the decoupled equation for $x$ is
redundant since the equation for $z$ is equivalent. In the massless
case $m=0$ we have $z=0$; hence, although the equation for $x$ does
not contribute to the dynamics, $x$ is needed in order to
reconstruct the spatial metric from the new variables.

The dimensionless dynamical system~(\ref{eq}) together with the
constraint~\eqref{omega} describes the full dynamics of the
Einstein-Vlasov system of Bianchi type~I. In the massive case the state space
associated with this system is the space of the variables
$\{(\Sigma_i, s_i, z)\}$, i.e.,
\begin{equation}\label{statespace}
\mathcal{X} :=
\left\{(\Sigma_i, s_i,z)\:\big|\: \left(\Sigma^2 < 1\right)
\wedge \left(s_i > 0\right) \wedge \left(0< z < 1\right)\right\}\:,
\end{equation}
where the $s_i$ and $\Sigma_i$ are subject to the constraints
$\sum_k\, s_k=1\:,\,\sum_k\,\Sigma_k = 0$. (The inequalities for
$s_i$ and $\Sigma_i$ follow from the definition~\eqref{defdimless}
and the constraint~\eqref{omega}, respectively.) The state space
$\mathcal{X}$ is thus five-dimensional.

It will turn out eventually that all solutions asymptotically
approach the boundaries of $\mathcal{X}$: $z=0$, $z=1$, $s_i=0$,
$\Omega = 0$ ($\Leftrightarrow \Sigma^2 =1$). This suggests to
include these sets in the analysis, whereby we obtain a compact
state space $\bar{\mathcal{X}}$.

The equations on the invariant subset $z=0$ of $\bar{\mathcal{X}}$
are identical to the coupled dimensionless system in the case of
massless particles $m=0$. We will therefore refer to the subset
$z=0$ as the massless subset; it represents the four-dimensional
state space for the massless case.

We conclude this section by looking at some variables in more
detail. The inequality $\Sigma^2 \leq 1$ together with the
constraint $\sum_k \Sigma_k =0$ results in $|\Sigma_i| \leq 2$ for
all $i$. Note that equality is achieved when
$(\Sigma_1,\Sigma_2,\Sigma_3) = (\pm 2 ,\mp 1, \mp 1)$ and
permutations thereof, cf.~Figure~\ref{kasneri}. The matter quantities
satisfy
\begin{equation}\label{wrelations}
0 \leq w \leq \textfrac{1}{3} \:, \qquad 0 \leq w_i \leq 3 w \leq 1\:.
\end{equation}
The equalities hold at the boundaries of the state space: $w =0$ iff
$z=1$, and $w = \textfrac{1}{3}$ iff $z=0$; $w_i = 0$ iff $s_i = 0$,
and $w_i = 3 w$ iff $s_i =1$ (provided that $z<1$; for $z=1$, $0 =
w_i = 3 w = 0$).

There exists a number of useful auxiliary equations that complement the system~\eqref{eq}:
\begin{align}
\label{omegaeq}
\Omega^\prime & =
\Omega\, \left[\,3(1-w)\Sigma^2 - \sum_k w_k\,\Sigma_k\,\right] \:,\\
\label{rhoeq} \rho' & =  -\rho\, [\,3(1+w) + \sum_k w_k\,\Sigma_k\,]
\leq -2\rho\:.
\end{align}
The inequality in~\eqref{rhoeq} follows by using $\Sigma_i \geq -2$
$\forall i$ and~\eqref{wrelations}. This shows that $\rho$ increases
monotonically toward the past which yields a matter singularity,
i.e., $\rho\rightarrow\infty$ for $\tau\rightarrow -\infty$. It is
often beneficial to consider the equations of the original variables
as auxiliary equations, e.g., $(g^{ii})^\prime=  -2g^{ii}\,(1 +
\Sigma_i)$.

\section{Fixed points, invariant subsets, and monotone functions}
\label{invfixed}

\subsection{Fixed points}
\label{fixed}

The dynamical system~\eqref{eq} possesses a number of fixed points,
all residing on the boundaries of the state space;
see~Table~\ref{fixtab}.

\begin{table}[ht]
\begin{center}
\begin{tabular}{|c|c|}\hline
Fixed point set & defined by \\ \hline
FS$^1$ & $z=1$, $\Sigma_j = 0 \:\,\forall j$ \\
KC$_i^1$ & $z=1$, $\Sigma^2 =1$, $s_i=1$, $s_j = 0\:\,\forall j \neq
i$
\\ \hline
TS$_i$ & $0\leq z\leq 1$, $\Sigma_i = 2$, $\Sigma_j =-1 \:\,\forall
j \neq i$, $s_i=0$  \\ \hline
F$^0$ & $z=0$, $\Sigma_j = 0 \:\,\forall j$, $w_j = 1/3 \:\,\forall j$  \\
D$_i^0$ & $z=0$, $s_i = 0$, $\Sigma_i =-1$, $\Sigma_{j}=1/2 =w_j\:\,
\forall j\neq i$  \\
QL$_i^0$ & $z=0$, $\Sigma_i = -2$, $\Sigma_j = 1 \:\,\forall j \neq
i$,
$s_i=0$  \\
KC$_i^0$ & $z=0$, $\Sigma^2 =1$, $s_i=1$, $s_j = 0 \:\,\forall j
\neq i$  \\ \hline
\end{tabular}
\end{center}
\caption{The fixed point sets. The range of the index $i$ is always
$i=1\ldots 3$. The superscript denotes the value of $z$; the first
kernel letter describes the type of fixed point set; if there is no
second kernel letter the fixed point set is just a point; if there
is a second kernel letter this letter denotes the dimensionality and
character of the set --- S refers to surface, L stands for line, and
C for circle.} \label{fixtab}
\end{table}

\begin{figure}[Ht]
\begin{center}
\psfrag{Ti1}[cc][cc][1][0]{$\text{T}^0_{i1}$}
\psfrag{Ti2}[cc][cc][1][0]{$\text{T}^0_{i2}$}
\psfrag{Ti3}[cc][cc][1][0]{$\text{T}^0_{i3}$}
\psfrag{Qi1}[cc][cc][1][0]{$\text{Q}^0_{i1}$}
\psfrag{Qi2}[cc][cc][1][0]{$\text{Q}^0_{i2}$}
\psfrag{Qi3}[cc][cc][1][0]{$\text{Q}^0_{i3}$}
\psfrag{S1}[cc][cc][1.2][0]{$\Sigma_1$}
\psfrag{S2}[cc][cc][1.2][0]{$\Sigma_2$}
\psfrag{S3}[cc][cc][1.2][0]{$\Sigma_3$}
\psfrag{S1m}[cc][cc][1][-90]{$\Sigma_1=-1$}
\psfrag{S2m}[cc][cc][1][30]{$\Sigma_2=-1$}
\psfrag{S3m}[cc][cc][1][-30]{$\Sigma_3=-1$}
\psfrag{S1p}[cc][cc][0.6][90]{$\Sigma_1=1$}
\psfrag{S2p}[cc][cc][0.6][30]{$\Sigma_2=1$}
\psfrag{S3p}[cc][cc][0.6][-30]{$\Sigma_3=1$}
\psfrag{s1}[cc][cc][1.0][90]{$\longleftarrow\: s_1 =0\:
\longrightarrow$} \psfrag{s2}[cc][cc][1.0][35]{$\longleftarrow \:s_2
=0 \:\longrightarrow$} \psfrag{s3}[cc][cc][1.0][-35]{$\longleftarrow
\:s_3 =0 \:\longrightarrow$}
\includegraphics[width=0.6\textwidth]{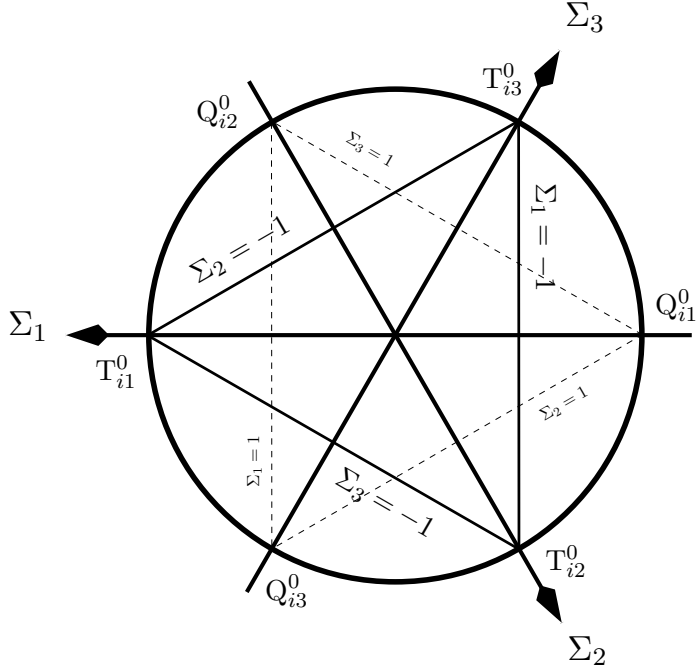}
\caption{The disc $\Sigma^2\leq 1$ and the Kasner circle $\text{KC}_i^0$.} \label{kasneri}
\end{center}
\end{figure}

$\text{FS}^1$ is a surface of fixed points that correspond to the
flat isotropic dust solution. The circles $\text{KC}_i^{1,0}$
consist of fixed points with constant $\Sigma_ i$, satisfying
$\Sigma^2=1,\sum_k\,\Sigma_k=0$, see Figure~\ref{kasneri}; these fixed
points correspond to Kasner solutions. The fixed points on
$\text{TS}_i$ are associated with the Taub representation of the
flat Minkowski spacetime. The intersection of $\text{TS}_i$ with
$(z=0)$ yields a line of fixed points which we denote
$\text{TL}_i^0$. The fixed points on $\text{QL}_i^0$ correspond to
the non-flat LRS Kasner solutions.
$\text{F}^0$ is a fixed point that corresponds to the flat isotropic
radiation solution. In Appendix~\ref{FRWLRS} we prove that $\text{F}^0$
is well-defined through the equations $w_1 = w_2= w_3 = 1/3$ (which
are to be solved for $(s_1,s_2,s_3)\,$). The location of
$\text{F}^0$ depends on the chosen distribution function, since the
equations $w_1 = w_2= w_3 = 1/3$ involve $f_0$.
Analogously, the equations $w_j = 1/2$ ($\forall \: j\neq i$) yield
a unique solution, the fixed point D$_i^0$; the location of the
point D$_i^0$ also depends on $f_0$. The fixed points D$_i^0$ are
associated with a scale-invariant LRS solution (related to a
distributional $f_0$; see Appendix~\ref{Si0space} for details).

\begin{figure}[Ht]
\begin{center}
\psfrag{TL1}[cc][cc][0.8][0]{$\text{TL}^0_1$}
\psfrag{TL2}[cc][cc][0.8][0]{$\text{TL}^0_2$}
\psfrag{TL3}[cc][cc][0.8][0]{$\text{TL}^0_3$}
\psfrag{QL1}[cc][cc][0.8][0]{$\text{QL}^0_1$}
\psfrag{QL2}[cc][cc][0.8][0]{$\text{QL}^0_2$}
\psfrag{QL3}[cc][cc][0.8][0]{$\text{QL}^0_3$}
\psfrag{T21}[cc][cc][0.6][0]{$\text{T}^0_{21}$}
\psfrag{T22}[cc][cc][0.6][0]{$\text{T}^0_{22}$}
\psfrag{T23}[cc][cc][0.6][0]{$\text{T}^0_{23}$}
\psfrag{Q21}[cc][cc][0.6][0]{$\text{Q}^0_{21}$}
\psfrag{Q22}[cc][cc][0.6][0]{$\text{Q}^0_{22}$}
\psfrag{Q23}[cc][cc][0.6][0]{$\text{Q}^0_{23}$}
\psfrag{T11}[cc][cc][0.6][0]{$\text{T}^0_{11}$}
\psfrag{T12}[cc][cc][0.6][0]{$\text{T}^0_{12}$}
\psfrag{T13}[cc][cc][0.6][0]{$\text{T}^0_{13}$}
\psfrag{Q11}[cc][cc][0.6][0]{$\text{Q}^0_{11}$}
\psfrag{Q12}[cc][cc][0.6][0]{$\text{Q}^0_{12}$}
\psfrag{Q13}[cc][cc][0.6][0]{$\text{Q}^0_{13}$}
\psfrag{T31}[cc][cc][0.6][0]{$\text{T}^0_{31}$}
\psfrag{T32}[cc][cc][0.6][0]{$\text{T}^0_{32}$}
\psfrag{T33}[cc][cc][0.6][0]{$\text{T}^0_{33}$}
\psfrag{Q31}[cc][cc][0.6][0]{$\text{Q}^0_{31}$}
\psfrag{Q32}[cc][cc][0.6][0]{$\text{Q}^0_{32}$}
\psfrag{Q33}[cc][cc][0.6][0]{$\text{Q}^0_{33}$}
\psfrag{KC1}[cc][cc][1.2][0]{$\text{KC}^0_1$}
\psfrag{KC2}[cc][cc][1.2][0]{$\text{KC}^0_2$}
\psfrag{KC3}[cc][cc][1.2][0]{$\text{KC}^0_3$}
\psfrag{s1}[cc][cc][1.0][90]{$\longleftarrow\: s_1 =0\:
\longrightarrow$} \psfrag{s2}[cc][cc][1.0][35]{$\longleftarrow \:s_2
=0 \:\longrightarrow$} \psfrag{s3}[cc][cc][1.0][-35]{$\longleftarrow
\:s_3 =0 \:\longrightarrow$}
\includegraphics[width=0.8\textwidth]{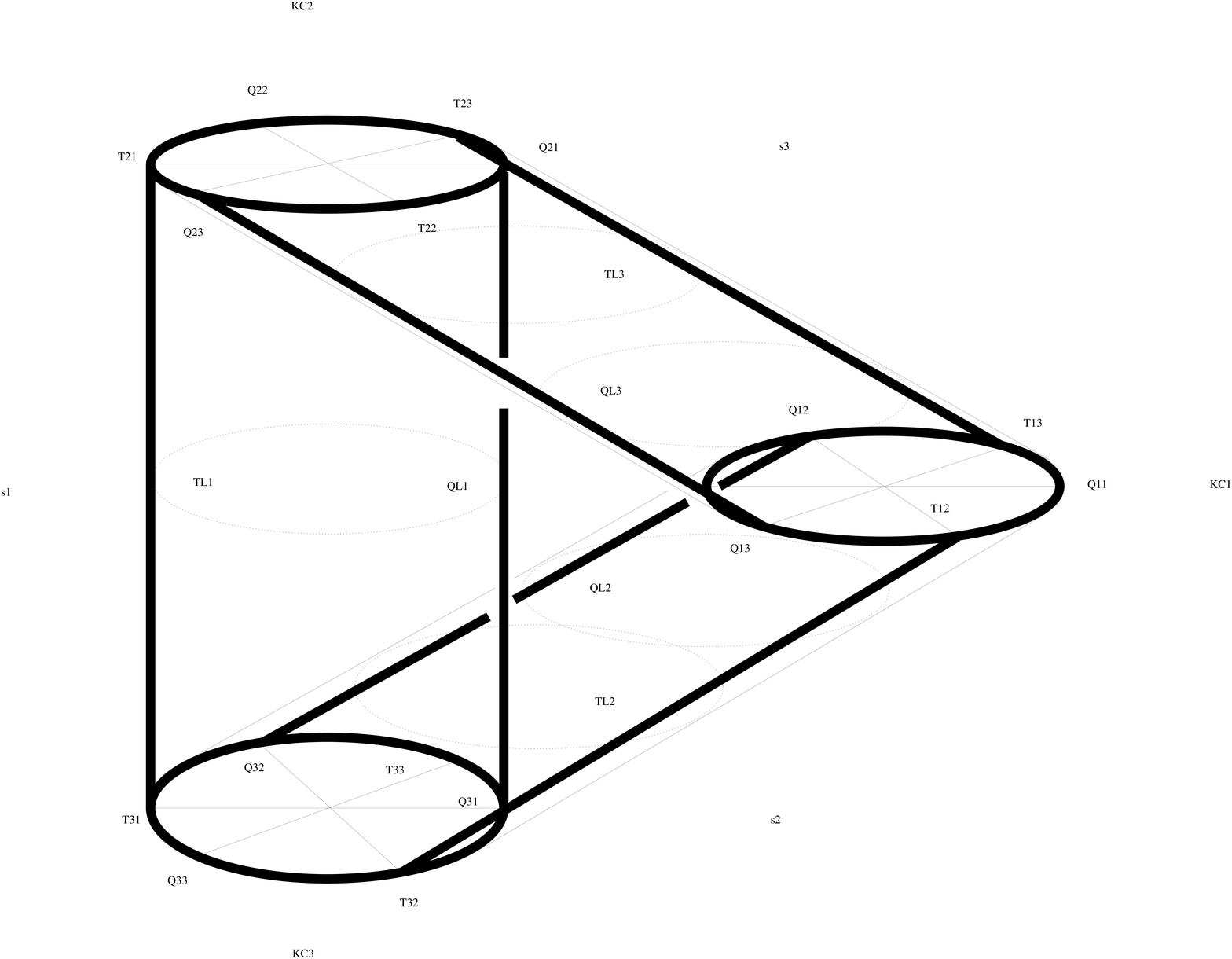}
\caption{A schematic depiction of the fixed points on $z=0$.
The underlying structure is the three sides of the $s_i$-triangle $s_1+s_2+s_3 =1$:
each point represents a disc $\Sigma^2\leq 1$; the vertices contain the Kasner circles $\text{KC}_i^0$.
Bold lines denote the lines of fixed points $\text{TL}_i^0$, $\text{QL}_i^0$,
and $\mathrm{KC}_i^0$.} \label{fixedp}
\end{center}
\end{figure}

The LRS points on $\text{KC}_i^0$ play a particularly important role
in the following, which motivates that they are given individual
names. We denote the three Taub points on KC$_i^0$ defined by
$\Sigma_j = 2$ (and thus $\Sigma_l = -1$ $\forall l\neq j$) by
$\text{T}_{ij}^0$, while we denote the three non-flat LRS point on
KC$_i^0$ given by $\Sigma_j = -2$ (and thus $\Sigma_l = 1$ $\forall
l\neq j$) by Q$_{ij}^0$. The Kasner circles $\text{KC}_j^0$ and
$\text{KC}_k^0$ are connected by the lines $\text{TL}_i^0$ and
$\text{QL}_i^0$; the end points of the line $\text{TL}_i^0$ are the
Taub points $\text{T}_{ji}^0$ and $\text{T}_{ki}^0$; analogously,
the end points of $\text{QL}_i^0$ are the points $\text{Q}_{ji}^0$
and $\text{Q}_{ki}^0$. (Here, $(i,j,k)$ is an arbitrary permutation
of $(1,2,3)$.) The remaining points $\text{T}_{ll}^0$ and
$\text{Q}_{ll}^0$ ($l=1\ldots 3$) do not lie on any of the fixed
point sets $\text{TL}^0_i$ or $\text{QL}^0_i$. This fixed point
structure is depicted in Figure~\ref{fixedp}.

\subsection{Invariant subsets and monotone functions}
\label{inv}

The dynamical system~\eqref{eq} possesses a hierarchy of invariant
subsets and monotone functions. Since this feature of the dynamical
system will turn out to be of crucial importance in the analysis of
the global dynamics, we give a detailed discussion.

$\mathcal{X}$: On the full (interior) state space $\mathcal{X}$ define
\begin{subequations}\label{m1eqs}
\begin{equation}\label{m1}
M_{(1)} = (s_1 s_2 s_3)^{-1/3} \frac{z}{1-z} \:.
\end{equation}
A straightforward computation shows
\begin{equation}
M_{(1)}^\prime = 2 M_{(1)}\:,
\end{equation}
\end{subequations}
i.e., $M_{(1)}$ is strictly monotonically increasing along orbits in
$\mathcal{X}$. Note that $M_{(1)}$ is intimately related to the
spatial volume density since $M_{(1)} = m^2 \det(g_{ij})^{1/3}$.

$\cZ^1$: This subset is characterized by $z=1$. Since $w_i = w =0$,
the equations for $s_i$ decouple, and the essential dynamics is
described by the equations $\Sigma_i^\prime = -(3/2)(1-\Sigma^2)
\Sigma_i$. (Note that these equations are identical to the Bianchi
type~I equations for dust --- it is therefore natural to refer to
$\cZ^1$ as the dust subset.) Explicit solutions for these equations
can be obtained by noting that
$\Sigma_1\propto\Sigma_2\propto\Sigma_3$ for all solutions, or by
using that $\Omega^\prime = 3\Sigma^2\,\Omega$.

$\cZ^0$: This subset is the massless boundary set $z=0$.
Since $w =1/3$, the dynamical system~\eqref{eq} reduces to
\begin{equation}\label{z0eq}
\Sigma_i^\prime = - \Omega\,[\,1+\Sigma_i -3 w_i\,]\:,\qquad
s_i^\prime = -2 s_i \,[\,\Sigma_i - \sum\nolimits_k s_k\, \Sigma_k \,]\:.
\end{equation}
Consider the function
\begin{subequations}\label{m2eqs}
\begin{equation}\label{m2}
M_{(2)} =\left(1 -\Sigma^2\right)^{-1} (s_1 s_2 s_3)^{-1/6} \int f_0
\left[ \sum\nolimits_k s_k v_k^2\right]^{1/2} d v_1 d v_2 d v_3\:.
\end{equation}
The derivative is
\begin{equation}
M_{(2)}^\prime = -2 \Sigma^2 M_{(2)}\:,
\end{equation}
which yields monotonicity when $\Sigma^2 \neq 0$. If $\Sigma^2 = 0$, then
\begin{equation}
M_{(2)}^\prime = 0 \:, \quad M_{(2)}^{\prime\prime} = 0 \:, \quad
M_{(2)}^{\prime\prime\prime} = -6 M_{(2)} \sum_i \left(w_i
-\textfrac{1}{3}\right)^2 \:.
\end{equation}
\end{subequations}
Hence, $M_{(2)}$ is strictly monotonically decreasing everywhere on
$z=0$, except at the fixed point $\text{F}^0$ (for which
$\Sigma^2=0$ and $w_1 = w_2 = w_3 = 1/3$), where $M_{(2)}$ attains a
(positive) minimum. The latter follows from the fact that
$(1-\Sigma^2)^{-1}$ is minimal at the point $\Sigma_i = 0
\:\,\forall i$ and that $\partial M_{(2)}/\partial s_i = (2
s_i)^{-1} [w_i - 1/3] M_{(2)}$.

$\cS_i$ ($i=1,2,3$): These invariant boundary subsets are defined by
$s_i=0$ (which yields $w_i=0$). There exists a monotone function on
$\mathcal{S}_1$,
\begin{equation}\label{m3}
M_{(3)}= (s_2 s_3)^{-1/2}\,\frac{z}{1-z}\, ,\qquad M_{(3)}^\prime =
(2-\Sigma_1)\,M_{(3)}\, ;
\end{equation}
analogous functions can be obtained on $\cS_2$ and $\cS_3$ through
permutations.

$\cK$: This boundary subset is the vacuum subset defined by
$\Omega=0$ (or equivalently $\Sigma^2=1$). The $\Sigma_i$ are
constant on this subset, which completely determines the dynamics of
the $s_i$ variables (via~\eqref{seq} or via the auxiliary equation
for $g^{ii}$). The Bianchi type~I vacuum solution is the familiar
Kasner solution and we thus refer to $\cK$ as the Kasner subset.

Intersections of the above boundary subsets yield boundary subsets
of lower dimensions; those that are relevant for the global dynamics
are discussed in the following.

$\cS_i^0$ and $\cS_i^1$: The intersection between the subset $\cS_i$
and $\cZ^0$ and $\cZ^1$ yields three-dimensional invariant subsets
$(s_i=0) \cap (z=0)$ and $(s_i=0) \cap (z=1)$, respectively. On
$\cS_i^0$ there exists a monotonically decreasing function:
\begin{equation}\label{m4}
M_{(4)} = (1+\Sigma_i)^2\:, \qquad M_{(4)}^\prime = - 2 \Omega M_{(4)} \:.
\end{equation}

$\cS_{ij}$: These subsets are defined by setting $s_i =0$ and $s_j
=0$ ($j\neq i$), i.e., $\cS_{ij} = \cS_i \cap \cS_j$. On $\cS_{ij}$,
we have $s_k =1$ ($k \neq i,j$) and $w_k = 3 w$, because $w_i = w_j
=0$.

$\cD_i^0$: The subsets $\cS_i^0$ admit two-dimensional invariant
subsets $\cD_i^0$ characterized by $(z=0) \cap (s_i=0) \cap (\Sigma_i =-1)$.
On $\cD_1^0$ consider the function
\begin{subequations}\label{m5eqs}
\begin{equation}
M_{(5)} = \left(2 + \Sigma_2 \Sigma_3 \right)^{-1} (s_2 s_3)^{-1/4}
\int f_0 \left[ s_2 v_2^2 + s_3 v_3^2\right]^{1/2} d v_1 d v_2 d
v_3\:;
\end{equation}
analogous functions can be defined on $\cD_2^0$ and $\cD_3^0$.
Eqs.~\eqref{z0eq} imply
\begin{equation}\label{m5d}
M_{(5)}^\prime  = -\textfrac{1}{12}\, M_{(5)} \left[ \left(1 -
2\Sigma_2\right)^2 + \left(1 - 2\Sigma_3\right)^2\right] \,,
\end{equation}
i.e., $M_{(5)}$ is strictly monotonically decreasing unless
$\Sigma_2 = 1/2 = \Sigma_3$. In the special case $\Sigma_2 = 1/2 =
\Sigma_3$ we obtain
\begin{equation}
M_{(5)}' = 0 \:, \quad M_{(5)}^{\prime\prime} = 0 \:, \quad
M_{(5)}^{\prime\prime\prime} =
-\textfrac{27}{8} M_{(5)} \left[ \left(w_2 -
\textfrac{1}{2}\right)^2 +\left(w_3 -
\textfrac{1}{2}\right)^2\right] \:.
\end{equation}
\end{subequations}
Hence, $M_{(5)}$ is strictly monotonically decreasing everywhere on
$\cD_1^0$ except for at the fixed point $\text{D}_1$, for which
$\Sigma_2=\Sigma_3=w_2=w_3=\frac{1}{2}$, cf.~Table~\ref{fixtab}. The
function $M_{(5)}$ possesses a positive minimum at $\text{D}_1$.
This is because $(2+\Sigma_2 \Sigma_3)^{-1}$ is minimal at the point
$\Sigma_2 = \Sigma_3 = 1/2$ and $\partial M_{(5)}/\partial s_i = (2
s_i)^{-1} [w_i - 1/2] M_{(5)}$ for $i=2,3$.

$\cK^0$: The intersection of the Kasner subset $\cK = (\Sigma^2 =1)$
with the $z=0$ subset yields a 3-dimensional subset, $\cK^0$. This
subset will play a prominent role in the analysis of the past
asymptotic behaviour of solution.

The remaining subsets we consider are not located at the boundaries
of the state space, but in the interior; these subsets are invariant
under the flow of the dynamical system, if the distribution function
$f_0$ satisfies certain symmetry conditions.

$\text{LRS}_i$: We define the subset $\text{LRS}_1$ of $\mathcal{X}$
through the equations $\Sigma_2=\Sigma_3$, $w_2=w_3$;
$\text{LRS}_{2,3}$ are defined analogously. In order for these sets
to be invariant under the flow of the dynamical system, the
distribution function $f_0$ must satisfy conditions that ensure
compatibility with the LRS symmetry, see Appendix~\ref{FRWLRS} for
details. For an orbit lying on $\text{LRS}_1$, Equation~(\ref{seq})
entails that $s_2(\tau) \propto s_3(\tau)$ (where the
proportionality constant exhibits a dependence on $f_0$, which
enters through the equation $w_2 =w_3$), and hence $g_{22}\propto
g_{33}$; by rescaling the coordinates one can achieve
$g_{22}=g_{33}$, i.e., a line element in an explicit LRS form.
Hence, the $\text{LRS}_i$ subsets, if present as invariant subsets,
comprise the solutions with LRS geometry.

FRW: If $f_0$ is compatible with an isotropic geometry, see
Appendix~\ref{FRWLRS} for details, the one-dimensional subset
characterized by the equations $\Sigma_i=0$ $\forall i$ and $w_1
=w_2=w_3 =w$ is an invariant subset (in fact: orbit), the FRW
subset. The equations $\Sigma_i=0$ yield $s_{i}=\mathrm{const}$,
whereby we obtain a Friedmann-Robertson-Walker (FRW) geometry, since
the spatial coordinates can be rescaled so that
$g_{ij}\propto\delta_{ij}$. Note that the location in $s_i$ of the
FRW subset depends on $f_0$, since the equations $w_1 = w_2= w_3$,
which are to be solved for $(s_1,s_2,s_3)$, involve $f_0$,
cf.~Appendix~\ref{FRWLRS}. Remarkably, in the massless case the
existence of a FRW solution (which corresponds to the fixed point
$\text{F}^0$) does not require any symmetry conditions on $f_0$.

\section{Local and global dynamics}
\label{locglo}

\subsection{Local dynamics}

Let us consider smooth reflection-symmetric
Bianchi type~I Vlasov solutions that approach fixed point sets when
$\tau\rightarrow -\infty$.

\begin{theorem}\label{locthm}
In the massive (massless) case there exists
\begin{itemize}
\item[(a)] a single orbit that approaches (corresponds
to) $\mathrm{F}^0$,
\item[(b)] three equivalent one-parameter sets of orbits (three single orbits)
that approach $\mathrm{D}_i^0$, $i=1 \ldots 3$,
\item[(c)] one three-parameter
(two-parameter) set of orbits that approaches $\mathrm{QL}_1^0$;
$\mathrm{QL}_2^0$ and $\mathrm{QL}_3^0$ yield equivalent sets,
\item[(d)] one four-parameter (three-parameter) set of orbits that
approaches the part of {\rm KC}$_1^0$ defined by $1<\Sigma_1<2$;
{\rm KC}$_2^0$ and {\rm KC}$_3^0$ yield equivalent sets.
\end{itemize}
\end{theorem}

\proof The statements of the theorem follow from the local stability
analysis of the fixed point sets F$^0$, D$_i^0$, $\text{QL}_i^0$,
$\text{KC}_i^0$, when combined with the Hartman-Grobman and the
reduction theorem, since the fixed points F$^0$, D$_i^0$ are
hyperbolic and $\text{QL}_i^0$, $\text{KC}_i^0$ are transversally
hyperbolic. This requires the dynamical system to be $\mathcal{C}^1$
and this leads to some restrictions on $f_0$. However, it is
possible to obtain an alternative proof that does not require such
restrictions. Such a proof can be obtained from the hierarchical
structure of invariant sets; we will refrain from making the details
explicit here, since our analysis of the global dynamics below
contains all essential ingredients implicitly.

\textit{Interpretation of Theorem~\ref{locthm} (massive case)}:
A three-parameter set of solutions 
converges to every non-LRS Kasner solution as $t\rightarrow 0$.
(In the state space description three equivalent sets of orbits approach three equivalent
transversally stable Kasner arcs that cover all non-LRS Kasner
solutions; the 
equivalence reflects the freedom of permuting the coordinates.)
Furthermore, a three-parameter set of
solutions approaches the non-flat LRS Kasner solution. 
Hence, in total, a four-parameter set of solutions asymptotically approaches 
non-flat Kasner states. 
There exist special solutions with
non-Kasner behaviour toward the singularity: one solution
isotropizes toward the singularity and a one-parameter set of solutions
approaches a non-Kasner LRS solution of the type~\eqref{Disol} 
(three equivalent one-parameter
sets of orbits approach three equivalent non-Kasner LRS fixed points
associated with this solution).
For the latter solutions 
$\Omega=3/4$; these solutions cannot be interpreted as a perfect
fluid solutions since they possess anisotropic pressures.

In the following we show that the list of Theorem~\ref{locthm} is
almost complete: there exist no other attractors toward the singularity
with one exception, a heteroclinic network that connects the flat LRS-Kasner
points.

\subsection{Global dynamics}
\label{globaldynamics}

\begin{theorem}\label{futurethm}
All orbits in the interior of the state space $\mathcal{X}$ of
massive particles [state space $\mathcal{Z}^0$ of massless
particles] converge to {\rm FS}$^1$ [{\rm F}$^0$] when
$\tau\rightarrow +\infty$; i.e., all smooth reflection-symmetric
Bianchi type~I Vlasov solutions isotropize toward the future.
\end{theorem}

A proof of this theorem has been given in~\cite{ren96}.
In Appendix~\ref{futureproof} we present an alternative proof based on dynamical
systems techniques.

\begin{theorem}\label{alphathm}
The $\alpha$-limit set of an orbit in the interior of the state
space is one of the fixed points $\mathrm{F}^0$, $\mathrm{D}_i^0$,
$\mathrm{QL}_i^0$, $\mathrm{KC}_i^0$, see Theorem~\ref{locthm}, or
it is the heteroclinic network $\cH^0$. The $\alpha$-limit set of a
generic orbit resides on the union of the fixed point sets {\rm
KC}$_i^0$ and possibly the heteroclinic network $\cH^0$.
\end{theorem}

\begin{remark}
A heteroclinic network is defined as a compact connected
flow-invariant set that is indecomposable (all points are connected
by pseudo-orbits), has a finite nodal set (the set of recurrent
points is a finite union of disjoint compact connected
flow-invariant subsets), and has finite depth (taking the
$\alpha$/$\omega$-limit iteratively yields the set of recurrent
points after a finite number of iterative steps); for details
see~\cite{Ashwin/Field:1999} and references therein. A simple case
is a heteroclinic network of depth $1$ whose nodes are equilibrium
points: it can be regarded as a collection of entangled heteroclinic
cycles. The heteroclinic network $\cH^0$ is of the latter type; it
will be introduced in the proof of the theorem.
\end{remark}

The remainder of this section is concerned with the proof of Theorem~\ref{alphathm}.
The first step in the proof is to gain a detailed understanding of the dynamics
on the relevant invariant subspaces of the dynamical system.

\subsubsection*{Dynamics on $\cS_i^0$}

\begin{lemma}\label{lemmaSi0}
Consider an orbit in the interior of $\cS_i^0$. Its $\alpha$-limit
set is either a fixed point on {\rm KC}$_j^0$ or {\rm KC}$_k^0$
($i\neq j \neq k$),  {\rm QL}$_i^0$ or {\rm TL}$_i^0$, or it is the
heteroclinic cycle $\cH_i^0$. The $\omega$-limit set consists of the
fixed point {\rm D}$_i^0$.
\end{lemma}

\begin{remark}
The heteroclinic cycle $\cH_i^0$ will be defined
in~\eqref{heterocycle}.
\end{remark}

\begin{figure}[Ht]
\begin{center}
\psfrag{A}[cc][cc][1.2][0]{$\text{TL}^0_1$}
\psfrag{B}[cc][rc][1.2][0]{$\text{QL}^0_1$}
\psfrag{K2}[cc][cc][1.2][0]{$\text{KC}^0_2$}
\psfrag{K3}[cc][cc][1.2][0]{$\text{KC}^0_3$}
\psfrag{T2}[cc][cc][0.9][0]{$\text{T}_{32}^0$}
\psfrag{T3}[cc][cc][0.9][0]{$\text{T}_{23}^0$}
\psfrag{P2}[cc][cc][0.9][0]{$\text{T}_{22}^0$}
\psfrag{P3}[cc][cc][0.9][0]{$\text{T}_{33}^0$}
\psfrag{S1}[cc][cc][0.8][0]{$\Sigma_1$}
\psfrag{S2}[cc][cc][0.8][0]{$\Sigma_2$}
\psfrag{S3}[cc][cc][0.8][0]{$\Sigma_3$}
\psfrag{Si}[cc][cc][0.8][0]{$\Sigma_i$}
\psfrag{s2}[cc][cc][0.8][0]{$s_2$}
\includegraphics[width=0.9\textwidth]{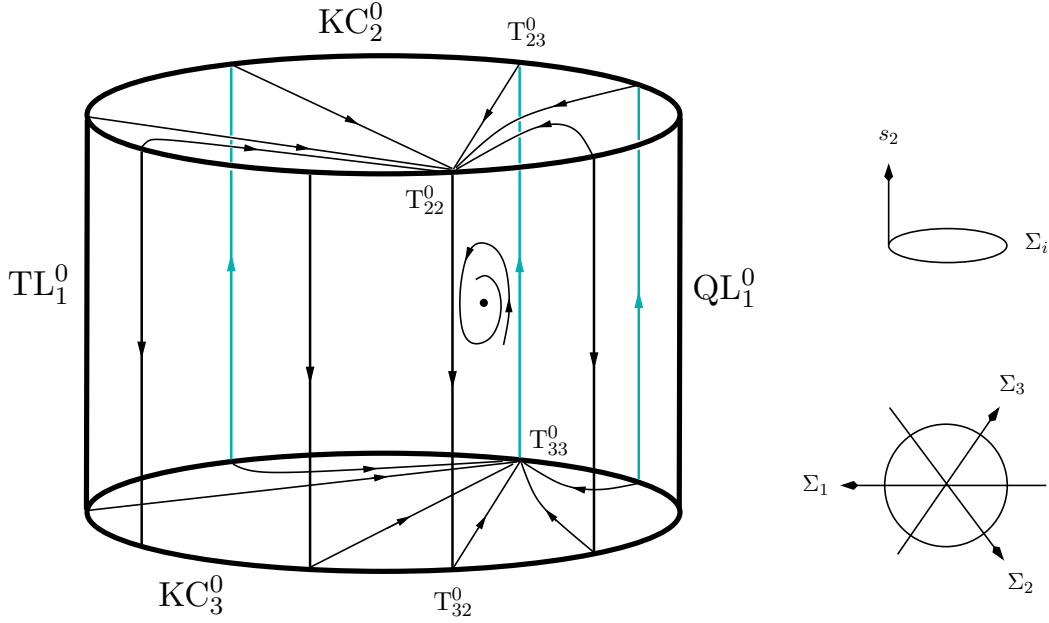}
\caption{Flow on the boundaries and on the invariant subset
$\Sigma_1 = -1$ on $\cS_1^0$. The fixed point on $\Sigma_1=-1$ is
the point D$_1^0$; the heteroclinic cycle $\cH_1^0$ consists of the
fixed points $\mathrm{T}_{22}^0$, $\mathrm{T}_{32}^0$,
$\mathrm{T}_{33}^0$, $\mathrm{T}_{23}^0$, and the connecting
orbits.}
\label{cylinder}
\end{center}
\end{figure}

\proof Without loss of generality we consider $\cS_1^0$, which can
be described by the variables
\begin{equation}
0< s_2 < 1 \;\, (s_3 = 1- s_2)
\quad\text{ and }\quad \Sigma_1, \Sigma_2,
\Sigma_3 \quad\left( \sum\nolimits_i \Sigma_i = 0,
\Sigma^2 < 1\right) \:;
\end{equation}
hence $\cS_1^0$ is represented by the interior of a cylinder, cf.~Figure~\ref{cylinder}.
The boundary of $\cS_1^0$ consists of the lateral boundary $\cS_1^0 \cap \cK^0$,
the base $\cS_{12}^0$, and the top surface $\cS_{13}^0$.

Since $\cS_1^0 \cap \cK^0$ is part of $\cK$, it follows that
$\Sigma_i \equiv \mathrm{const}$ for all orbits on $\cS_1^0 \cap \cK^0$.
We observe that $s_2$ is monotonically increasing
(decreasing) when $\Sigma_2 < \Sigma_3$ ($\Sigma_2 > \Sigma_3$), since
$s_2^\prime = -2 s_2 (1-s_2)(\Sigma_2 -\Sigma_3)$;
the two domains are separated by the lines of fixed points
TL$_1^0$ and QL$_1^0$, see Figure~\ref{cylinder}.

The key equations to understand the flow on $\cS_{12}^0$ are
\begin{equation}
\Omega^\prime = \Omega (2 \Sigma^2 - \Sigma_3)\quad\text{ and } \quad
\Sigma_3^\prime = \Omega (2-\Sigma_3)\, .
\end{equation}
From the first equation it follows that all points on
KC$_3^0$ are transversally hyperbolic repelling fixed points
except for T$_{33}^0$; from the second
equation we infer that T$_{33}^0$ is the attractor of the entire
interior of $\cS_{12}^0$.
Similarly, T$_{22}^0$ is the attractor on $\cS_{13}^0$, see Figure~\ref{cylinder}.

The plane $\cD_1^0$, defined by $\Sigma_1 = -1$, is an invariant subset in $\cS_1^0$.
In the interior of the plane we find the fixed point
D$_1^0$; the boundary consists of a heteroclinic cycle $\cH_1^0$,
\begin{equation}\label{heterocycle}
\cH_1^0: \mathrm{T}_{22}^0 \rightarrow \mathrm{T}_{32}^0 \rightarrow
\mathrm{T}_{33}^0 \rightarrow \mathrm{T}_{23}^0 \rightarrow
\mathrm{T}_{22}^0\, .
\end{equation}
(Note that analogous cycles $\cH_2^0$ and $\cH_3^0$ exist
on $\cS_2^0$ and $\cS_3^0$, respectively.)
The function $M_{(5)}$ is monotonically decreasing on $\cD_1^0$, cf.~\eqref{m5eqs}.
Application of the monotonicity principle, see Appendix~\ref{dynsys},
yields that $\text{D}_1^0$ is the $\omega$-limit and that $\cH_1^0$ is the
$\alpha$-limit for all orbits on $\cD_1^0$, cf.~Figure~\ref{cylinder}.

Consider now an orbit in $\cS_1^0$ with $\Sigma_1 \neq -1$.
The function $M_{(4)} = (1+\Sigma_1)^2$ is monotonically decreasing
on $\cS_1^0$, cf.~\eqref{m4}. The monotonicity
principle implies that the $\omega$-limit lies on
$\Sigma_1 = -1$ or $\Sigma^2 =1$ (but $\Sigma_1 \neq \pm 2$).
Since the logarithmic derivative of $\Omega$ is positive everywhere on $\cS_1^0 \cap \cK^0$
(except at $\text{T}^0_{22}$ and $\text{T}^0_{33}$), i.e.,
$\Omega^{-1}\Omega^\prime|_{\Omega =0} = 2  -\sum_k w_k \Sigma_k> 0$,
it follows that the ``wall'' $\cS_1^0 \cap \cK^0$ of the cylinder
is repelling everywhere away from $\Sigma_1 = -1$.
Consequently, the
$\omega$-limit of the orbit cannot lie on $\Sigma^2 =1$
but is contained in $\Sigma_1 = -1$.
The fixed point $\text{D}_1^0$ is a hyperbolic sink, as we conclude
from the dynamics on $\Sigma_1 = -1$ and from
$(1+\Sigma_1)^{-1} (1+ \Sigma_1)^\prime|_{\text{D}_1^0} = -3/4$.
Therefore, the a priori possible $\omega$-limit sets on $\Sigma_1 = -1$
are $\text{D}_1^0$ and the heteroclinic cycle $\cH_1^0$.

To prove that $\text{D}_1^0$ is the actual $\omega$-limit we again
consider the function $M_{(5)}$. However, we no longer restrict its
domain of definition to $\mathcal{D}_1^0$, but view it as a function
on $\cS_1^0$; we obtain
\begin{equation}\label{M52}
12 M_{(5)}^\prime = -M_{(5)} \left[(\Sigma_1 + 2 \Sigma_2)^2 +
(\Sigma_1+2 \Sigma_3)^2 + 6 (\Sigma_1+1)^2 - 6(\Sigma_1+1)\right]\:.
\end{equation}
The bracket is positive when $\Sigma_1 < -1$; hence $M_{(5)}$ is
decreasing when $\Sigma_1 < -1$. This prevents orbits with $\Sigma_1
< -1$ from approaching $\cH_1^0$, since the cycle is characterized
by $M_{(5)} = \infty$. Now suppose that there exist an orbit in
$\Sigma_1 > -1$, whose $\omega$-limit is $\cH_1^0$. At late times
the trajectory shadows the cycle; hence, for late times, the bracket
in~\eqref{M52} is almost always positive along the trajectory --
only when the trajectory passes through a small neighbourhood of
$(\Sigma_1,\Sigma_2,\Sigma_3) = (-1,1/2,1/2)$ the bracket is
marginally negative. Since the trajectory spends large amounts of
time near the fixed points and the passages from one fixed point to
another become shorter and shorter in proportion,
it follows that at late times $M_{(5)}$ is decreasing along the
orbit (with ever shorter periods of limited increase). This is a
contradiction to the assumption that the orbit is attracted by the
heteroclinic cycle. We therefore draw the conclusion that
$\text{D}_1^0$ is the global sink on $\cS_1^0$.

Consider again an orbit in $\cS_1^0$ with $\Sigma_1 \neq -1$.
Invoking the monotonicity principle with the function $M_{(4)}$ we
find that the $\alpha$-limit of the orbit must be located on
$\Sigma^2 =1$, $\Sigma_1 \neq -1$. From the analysis of the flow on
the boundaries of the cylinder we obtain that all fixed points on
$\Sigma^2=1$ except for T$_{22}^0$ and T$_{33}^0$ are transversally
hyperbolic. The fixed points on KC$_2^0$ with $\Sigma_2 < \Sigma_3$
and the points on KC$_3^0$ with $\Sigma_2
> \Sigma_3$ are saddles; the fixed points on KC$_2^0$ with
$\Sigma_2 > \Sigma_3$ and those on KC$_3^0$ with $\Sigma_2 <
\Sigma_3$ are transversally hyperbolic sources (except for
T$_{22}^0$, T$_{33}^0$): every point attracts a one-parameter set of
orbits from $\cS_1^0$ as $\tau\rightarrow -\infty$. In contrast,
each fixed point on TL$_1^0$ and QL$_1^0$ is a source for exactly
one orbit. The structure of the flow on $\Sigma^2=1$ implies that
the $\alpha$-limit of the orbit in $\cS_1^0$ with $\Sigma_1 \neq -1$
must be one of the transversally hyperbolic sources.

This establishes Lemma~\ref{lemmaSi0}.

\subsubsection*{Dynamics on $\cK^0$}

The invariant subset $\cK^0$ is defined by setting $z=0$ and $\Sigma^2=1$;
it can be represented by the Cartesian
product of the circle ($\Sigma^2=1$) in the $\Sigma_i$-space times
the $s_i$-triangle given by $\{0<s_1,s_2,s_3<1$, $\sum_k s_k = 1\}$. The flow
on this space possesses a simple structure: since $\Sigma_i^\prime
\equiv \mathrm{const}$ for all orbits, the dynamical freedom resides
in the $s_i$-spaces.

\begin{figure}[Ht]
\begin{center}
\psfrag{S1}[cc][cc][0.9][0]{$\Sigma_1$}
\psfrag{S2}[cc][cc][0.9][0]{$\Sigma_2$}
\psfrag{S3}[cc][cc][0.9][0]{$\Sigma_3$}
\psfrag{s1}[cc][cc][0.9][0]{$s_1$}
\psfrag{s2}[cc][cc][0.9][0]{$s_2$}
\psfrag{s3}[cc][cc][0.9][0]{$s_3$}
\psfrag{s10}[cc][cc][0.7][55]{$s_1=0$}
\psfrag{s20}[cc][cc][0.7][0]{$s_2=0$}
\psfrag{s30}[cc][cc][0.7][-60]{$s_3=0$}
\psfrag{K1}[lc][lc][0.9][0]{$\text{KC}^0_1$}
\psfrag{K2}[cc][cc][0.9][0]{$\text{KC}^0_2$}
\psfrag{K3}[rc][rc][0.9][0]{$\text{KC}^0_3$}
\psfrag{P}[cc][cc][0.7][0]{$\mathrm{P}$}
\includegraphics[width=0.9\textwidth]{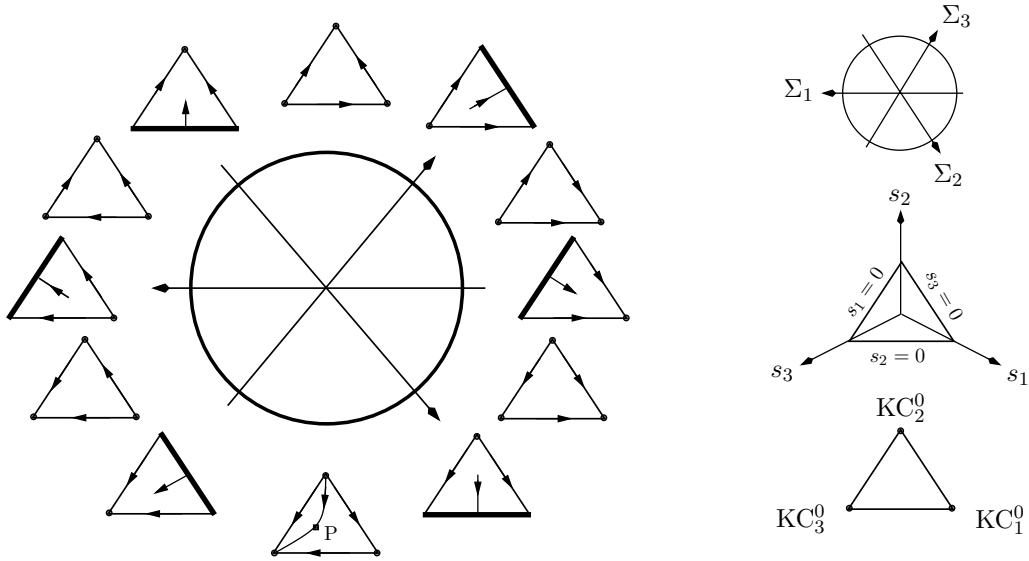}
\caption{Schematic representation of the flow of the dynamical
system on the invariant set $\cK^0 =(z=0) \cap (\Sigma^2 =1)$, which
is the Cartesian product of the $(\Sigma^2=1)$-circle and the
$s_i$-triangle. The depicted fixed points are the Kasner fixed
points and the lines of fixed points $\text{TL}_i^0$
($\leftrightarrow \Sigma_i =2$) and $\text{QL}_i^0$
($\leftrightarrow\Sigma_i = -2$). All orbits are heteroclinic, like
the orbit through the arbitrary point $\mathrm{P}$ that connects
$\text{KC}_2^0$ with $\text{KC}_3^0$.} \label{sigmaquadgleich1}
\end{center}
\end{figure}

A schematic depiction of the flow on $\cK^0$ is given in
Figure~\ref{sigmaquadgleich1}. All fixed points are located on the
boundaries of $\cK^0$, i.e., on $s_1=0$, $s_2=0$, or $s_3=0$. The
vertices of the $s_i$-triangle are the Kasner circles KC$_i^0$. If
$(\Sigma_1,\Sigma_2,\Sigma_3) \in (\Sigma^2 =1)$ is such that
$\Sigma_k = 2$ (respectively $\Sigma_k = -2$) for some $k$, then the
side $s_k = 0$ of the triangle is a line of fixed points, TL$_k^0$
(respectively QL$_k^0$). Note that all fixed points are
transversally hyperbolic on $\cK^0$, and that they constitute the
$\alpha$- and $\omega$-limit sets for all orbits on $\cK^0$. The
character of the fixed points, i.e., whether they are (transversal)
attractors or repellors, depends on the sector of the circle
($\Sigma^2=1$), see~Figure~\ref{sigmaquadgleich1}.

The results about the global dynamics on $\cS^0_i$ and $\cK^0$
will turn out to be an integral part in the proof of Theorem~\ref{alphathm},
which we will address next.
First, we will prove the massless case of the theorem.

\subsubsection*{Dynamics on $\cZ^0$}

Let $\gamma$ be an arbitrary orbit in the interior of $\cZ^0$,
$\gamma \neq \{ \text{F}^0 \}$. The function $M_{(2)}$ is strictly
monotonically decreasing on $\cZ^0$ (except at $\text{F}^0$, where
it has a minimum), cf.~\eqref{m2eqs}ff.; hence we can use the
monotonicity principle: the $\alpha$-limit set $\alpha(\gamma)$ of
$\gamma$ must be located on the boundaries of $\cZ^0$, i.e., on
$\cS_i^0$ or $\cK^0$. The first step in our analysis is to prove
that the interior of the subsets $\cS_i^0$ and $\cK^0$ cannot belong
to $\alpha(\gamma)$, unless $\gamma$ is one of three special orbits.

Recall from our analysis of $\cS_i^0$ that the fixed point
$\text{D}_i^0 \in \cS_i^0$ is a hyperbolic sink on $\cS_i^0$. In the
orthogonal direction, however, we obtain $s_i^{-1}s_i^\prime
|_{\text{D}_i^0} = 3$. It follows that $\text{D}_i^0$ is a
hyperbolic saddle in the state space $\cZ^0$ and that there exists
exactly one orbit $\delta^0_i$ that emanates from $\text{D}_i^0$
into the interior of $\cZ^0$. (Theorem~\ref{futurethm} implies that
$\delta^0_i$ converges to the global sink $\text{F}^0$ as
$\tau\rightarrow \infty$.)

Henceforth, let $\gamma$ be different from $\delta^0_i$. In order to
show that $\alpha(\gamma)$ does not contain any point of the
interior of $\cS^0_i$, we perform a proof by contradiction: assume
that $\alpha(\gamma)$ contains a point $\mathrm{P}$ of the interior
of $\cS_i^0$; then the whole orbit through $\mathrm{P}$ and the
$\omega$-limit $\omega(\mathrm{P})$ (as well as the $\alpha$-limit)
of that orbit must be contained in $\alpha(\gamma)$. As already
shown, $\text{D}_i^0$ is the global attractor on $\cS_i^0$, hence
$\alpha(\gamma) \ni \omega(\mathrm{P}) = \text{D}_i^0$. Since the
saddle $\text{D}_i^0$ is in $\alpha(\gamma)$, the unique orbit
$\delta^0_i$ emanating from it is contained in $\alpha(\gamma)$ as
well. Thus, ultimately, $\omega(\delta^0_i)$, i.e., the point
$\text{F}^0$, must be contained in $\alpha(\gamma)$; this is a
contradiction, since $\text{F}^0$ is a sink. Therefore, $\gamma$
cannot contain any $\alpha$-limit point in the interior of
$\cS_i^0$. We will now use similar reasoning repeatedly.

Assume next that $\alpha(\gamma)$ contains a point $\mathrm{P}$ in
the interior of $\cK^0$, i.e., a point with $\Sigma^2 =1$, $0< s_i
<1\:\forall i$. Suppose first that $\Sigma_k\neq \pm 2$ for all $k$.
Since $\mathrm{P}$ is an element of $\alpha(\gamma)$, the whole
orbit through $\mathrm{P}$ and the $\alpha$-limit
$\alpha(\mathrm{P})$ of that orbit must be contained in
$\alpha(\gamma)$. From the dynamics on $\cK^0$,
cf.~Figure~\ref{sigmaquadgleich1}, it follows that
$\alpha(\mathrm{P})$ is one of the Kasner fixed points on
$\text{KC}_i^0$, where $i$ corresponds to the direction determined
by $\Sigma_i = \max_k \Sigma_k$; we hence denote
$\alpha(\mathrm{P})$ as $\mathrm{K}_\mathrm{P}$. Since $\Sigma_k\neq
\pm 2$, it follows from the previous analysis that
$\mathrm{K}_\mathrm{P}$ is a transversally hyperbolic source on the
subspace $\cK^0$; $\Omega^{-1} \Omega^\prime|_{\Sigma^2=1} = 2 -
\sum_k w_k \Sigma_k > 0$ yields that $\mathrm{K}_\mathrm{P}$ is a
transversally hyperbolic source on the whole space $\cZ^0$. Since
$\alpha(\gamma)$ contains the transversally hyperbolic source
$\mathrm{K}_\mathrm{P}$, that fixed point necessarily constitutes
the entire $\alpha$-limit set, i.e., $\alpha(\gamma)
=\mathrm{K}_\mathrm{P}$. This is in contradiction to our assumption
$\alpha(\gamma) \ni \mathrm{P}$. The omitted cases $\Sigma_i = \pm
2$ for some $i$ will be dealt with next.

Suppose that $\Sigma_i = -2$ for one index $i$. Assume that
$\mathrm{P}$ lies in $\alpha(\gamma)$, therefore
$\alpha(\mathrm{P})$ is contained in $\alpha(\gamma)$ as well. The
dynamics on $\cK^0$ implies that $\alpha(\mathrm{P})$ is a fixed
point $\mathrm{Q}_{\mathrm{P}}$ on QL$_i^0$,
cf.~Figure~\ref{sigmaquadgleich1}. This point is a transversally
hyperbolic source; $\Omega^{-1} \Omega^\prime|_{\text{QL}_i^0} = 1$
in this case. By the same argument as above we obtain a
contradiction to the assumption $\alpha(\gamma) \ni \mathrm{P}$.

Finally suppose that $\Sigma_i = 2$ for one index $i$. When we
assume that $\mathrm{P}$ is in $\alpha(\gamma)$, then the
$\omega$-limit $\omega(\mathrm{P})$ is contained in
$\alpha(\gamma)$. From Figure~\ref{sigmaquadgleich1} we see that
$\omega(\mathrm{P})$ is a fixed point $\mathrm{T}_{\mathrm{P}}$ on
TL$_i^0$. The point $\mathrm{T}_{\mathrm{P}}$ is a transversally
hyperbolic saddle, since $\Omega^{-1} \Omega^\prime|_{\text{TL}_i^0}
= 3$, and there exists exactly one orbit that emanates from it,
namely the orbit that connects $\mathrm{T}_{\mathrm{P}}$ with
$\text{D}_i$ in $\cS_i^0$. Since $\mathrm{T}_{\mathrm{P}} \in
\alpha(\gamma)$, that orbit must also be contained in
$\alpha(\gamma)$. This is in contradiction to the previous result:
$\alpha(\gamma)$ cannot contain interior points of $\cS_i^0$. Hence
our assumption $\alpha(\gamma) \ni \mathrm{P}$ was false: the
$\alpha$-limit of $\gamma$ cannot contain any interior point of
$\cK^0$.

Our analysis results in the following statement: There exist four
special orbits, one trivial orbit corresponding to the fixed point
$\text{F}^0$ and three orbits, the orbits $\delta^0_i$, that
converge to the fixed points $\text{D}^0_i \in \cD^0_i$. The
$\alpha$-limit set of every orbit $\gamma$ in $\cZ^0$ different from
$\text{F}^0$ and $\delta^0_i$ must be located on the boundaries of
the spaces $\cS_i^0$ and $\cK^0$, i.e., on the union of the
boundaries of the cylinders $\cS_i^0$, which we denote by
$\partial\cS^0=\partial \cS_1^0 \cup
\partial \cS_2^0 \cup
\partial \cS_3^0$.
The set $\partial\cS^0$ is depicted in Figure~\ref{fixedp}: it
comprises the lateral surfaces of the cylinders and the base/top
surfaces.

All fixed points on $\partial\cS^0$ are transversally hyperbolic except
for the points $\text{T}_{ii}^0$:
$\text{TL}_i^0$ consists of transversally hyperbolic saddles; in
contrast, the fixed points on $\text{QL}_i^0$ are transversally
hyperbolic sources; points on $\text{KC}_i^0$ with $\Sigma_i
> 1$, $\Sigma_i \neq 2$ are sources while those with $\Sigma_i < 1$
are saddles.
Combining the analysis of the preceding sections,
see~Figs.~\ref{cylinder} and~\ref{sigmaquadgleich1}, we obtain, more
specifically: each point on $\text{QL}_i^0$ is a source for a
one-parameter family of orbits that emanate into the interior of
$\mathcal{Z}^0$, and each point on $\text{KC}_i^0$ with $\Sigma_i >
1$ ($\Sigma_i \neq 2$) is the source for a two-parameter family.
(The points with $\Sigma_i =1$ on $\text{KC}_i^0$ are the two points
$\text{Q}^0_{ij} \in \text{QL}_j^0$ and $\text{Q}^0_{ik}\in
\text{QL}_k^0$. Each of these two points is a transversally
hyperbolic source for a one-parameter family of orbits, however,
those orbits are not interior orbits, but remain on the boundary of
$\mathcal{Z}^0$.)

\begin{figure}[Ht]
\begin{center}
\psfrag{TL1}[cc][cc][0.8][0]{$\text{TL}^0_1$}
\psfrag{TL2}[cc][cc][0.8][0]{$\text{TL}^0_2$}
\psfrag{TL3}[cc][cc][0.8][0]{$\text{TL}^0_3$}
\psfrag{QL1}[cc][cc][0.8][0]{$\text{QL}^0_1$}
\psfrag{QL2}[cc][cc][0.8][0]{$\text{QL}^0_2$}
\psfrag{QL3}[cc][cc][0.8][0]{$\text{QL}^0_3$}
\psfrag{T21}[cc][cc][0.6][0]{$\text{T}^0_{21}$}
\psfrag{T22}[cc][cc][0.6][0]{$\text{T}^0_{22}$}
\psfrag{T23}[cc][cc][0.6][0]{$\text{T}^0_{23}$}
\psfrag{Q21}[cc][cc][0.6][0]{$\text{Q}^0_{21}$}
\psfrag{Q22}[cc][cc][0.6][0]{$\text{Q}^0_{22}$}
\psfrag{Q23}[cc][cc][0.6][0]{$\text{Q}^0_{23}$}
\psfrag{T11}[cc][cc][0.6][0]{$\text{T}^0_{11}$}
\psfrag{T12}[cc][cc][0.6][0]{$\text{T}^0_{12}$}
\psfrag{T13}[cc][cc][0.6][0]{$\text{T}^0_{13}$}
\psfrag{Q11}[cc][cc][0.6][0]{$\text{Q}^0_{11}$}
\psfrag{Q12}[cc][cc][0.6][0]{$\text{Q}^0_{12}$}
\psfrag{Q13}[cc][cc][0.6][0]{$\text{Q}^0_{13}$}
\psfrag{T31}[cc][cc][0.6][0]{$\text{T}^0_{31}$}
\psfrag{T32}[cc][cc][0.6][0]{$\text{T}^0_{32}$}
\psfrag{T33}[cc][cc][0.6][0]{$\text{T}^0_{33}$}
\psfrag{Q31}[cc][cc][0.6][0]{$\text{Q}^0_{31}$}
\psfrag{Q32}[cc][cc][0.6][0]{$\text{Q}^0_{32}$}
\psfrag{Q33}[cc][cc][0.6][0]{$\text{Q}^0_{33}$}
\psfrag{KC1}[cc][cc][1.2][0]{$\text{KC}^0_1$}
\psfrag{KC2}[cc][cc][1.2][0]{$\text{KC}^0_2$}
\psfrag{KC3}[cc][cc][1.2][0]{$\text{KC}^0_3$}
\psfrag{s1}[cc][cc][1.0][90]{$\longleftarrow\: s_1 =0\: \longrightarrow$}
\psfrag{s2}[cc][cc][1.0][35]{$\longleftarrow \:s_2 =0 \:\longrightarrow$}
\psfrag{s3}[cc][cc][1.0][-35]{$\longleftarrow \:s_3 =0 \:\longrightarrow$}
\includegraphics[width=0.8\textwidth]{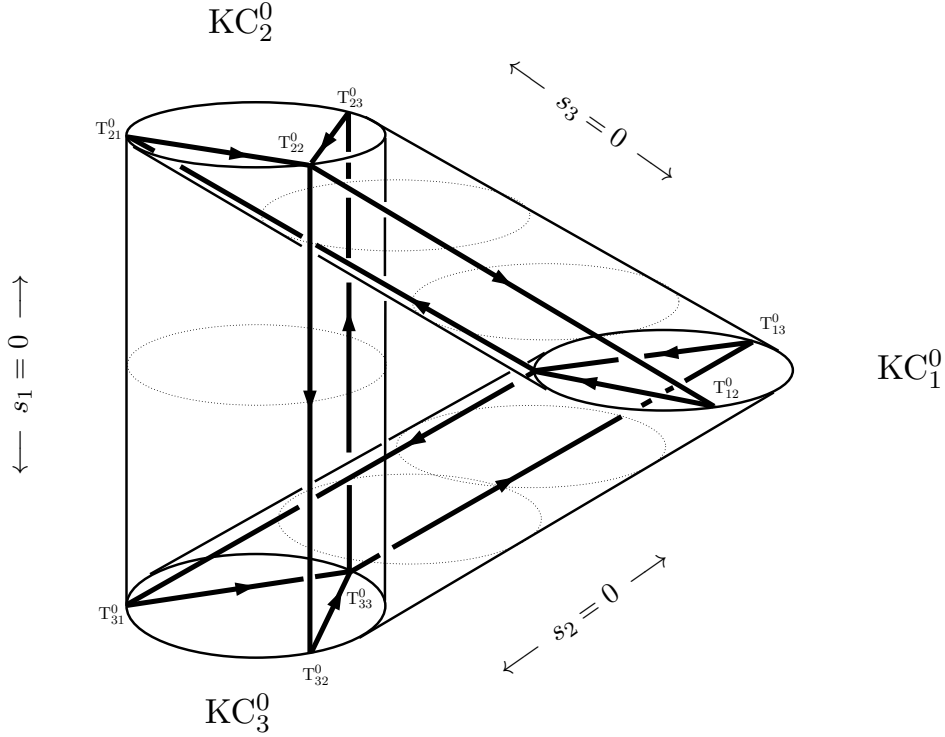}
\caption{The heteroclinic network $\cH^0$ that exists on the set
$\partial\cS^0$. Its building blocks are the heteroclinic cycles
$\cH^0_1$, $\cH^0_2$, $\cH^0_3$.} \label{network}
\end{center}
\end{figure}

\begin{figure}[Ht]
\begin{center}
\psfrag{T11}[cc][cc][0.9][0]{$\text{T}_{11}^0$}
\psfrag{T22}[cc][cc][0.9][0]{$\text{T}_{22}^0$}
\psfrag{T33}[cc][cc][0.9][0]{$\text{T}_{33}^0$}
\psfrag{T12}[cc][cc][0.9][0]{$\text{T}_{12}^0$}
\psfrag{T21}[cc][cc][0.9][0]{$\text{T}_{21}^0$}
\psfrag{T13}[cc][cc][0.9][0]{$\text{T}_{13}^0$}
\psfrag{T31}[cc][cc][0.9][0]{$\text{T}_{31}^0$}
\psfrag{T23}[cc][cc][0.9][0]{$\text{T}_{23}^0$}
\psfrag{T32}[cc][cc][0.9][0]{$\text{T}_{32}^0$}
\includegraphics[width=0.7\textwidth]{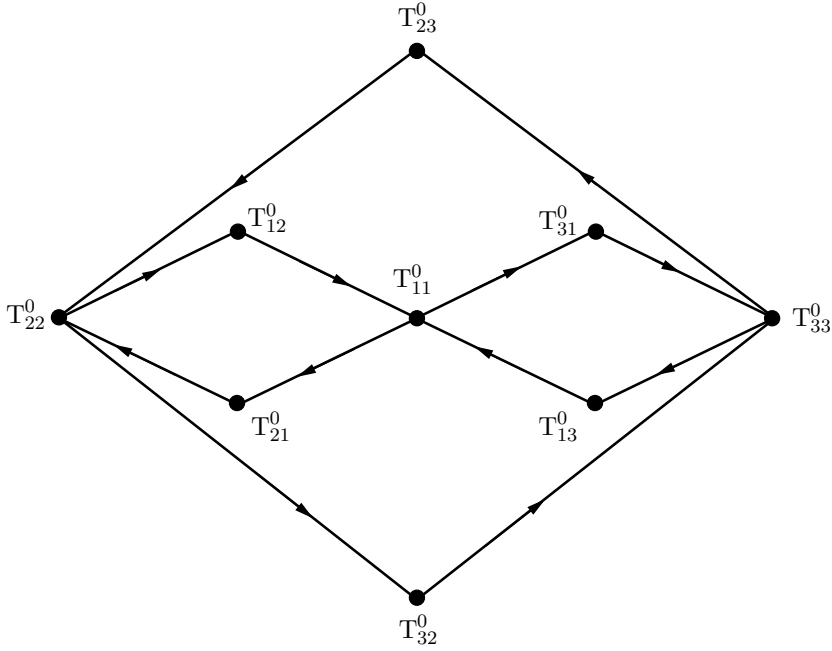}
\caption{Schematic representation of $\cH^0$.}
\label{networkschematic}
\end{center}
\end{figure}

The non-transversally hyperbolic fixed points $\text{T}_{ii}^0$ are
part of a special structure that is present on $\partial\cS^0$: the
set $\partial\cS^0$ exhibits a robust heteroclinic network $\cH^0$
(of depth $1$), see, e.g.,~\cite{Ashwin/Field:1999} for a discussion
of heteroclinic networks; the network $\cH^0$ is depicted in
Figure~\ref{network}; a schematic depiction is given in
Figure~\ref{networkschematic}. In particular we observe that the
heteroclinic cycles $\cH_i^0$ of the spaces $\cS_i^0$ appear as
heteroclinic subcycles of the network.

A straightforward analysis of the flow on $\partial \cS^0$ using the
same type of reasoning as above leads to the result that there exist
no other structures on $\partial \cS^0$ that could serve as
$\alpha$-limits for an interior $\mathcal{Z}^0$-orbit $\gamma$. We
have thus proved the following statement: The $\alpha$-limit of
$\gamma$ is one of the transversally hyperbolic sources listed
above, or it is the heteroclinic network (or a heteroclinic subcycle
thereof). This concludes the proof of the massless case of
Theorem~\ref{alphathm}.

\subsection*{Global dynamics in the massive case}

Let $\gamma$ be an arbitrary orbit in the interior of the state
space $\mathcal{X}$. The function $M_{(1)}$ is strictly
monotonically increasing on $\mathcal{X}$ (and on $\mathcal{K}$),
cf.~\eqref{m1eqs}ff.; moreover, $M_{(1)}$ vanishes for $z\rightarrow
1$ and $s_i \rightarrow 0$ (unless $z\rightarrow 0$ simultaneously).
Hence, by applying the monotonicity principle we obtain that the
$\alpha$-limit set $\alpha(\gamma)$ of $\gamma$ must be located on
$\cZ^0$ including its boundaries.

Consider the fixed point $\text{F}^0 \in \cZ^0$. By
Theorem~\ref{futurethm} this fixed point is a global sink on
$\cZ^0$. In the orthogonal direction, however, we have $z^{-1}
z^\prime |_{\text{F}^0} = 2$. It follows that $\text{F}^0$ is a
hyperbolic saddle in the space $\cX$ and that there exists exactly
one orbit $\phi$ that emanates from $\text{F}^0$ into the interior
of $\cX$. (Theorem~\ref{futurethm} implies that $\phi$ converges to
$\text{FS}^1$ as $\tau\rightarrow \infty$; thus, $\phi$ represents
the unique solution of the Einstein-Vlasov equations that
isotropizes toward the past and the future.)

Let $\gamma$ be different from $\phi$. Assume that $\alpha(\gamma)$
contains a point $\mathrm{P}$ of the interior of $\cZ^0$; then the
whole orbit through $\mathrm{P}$ and the $\omega$-limit
$\omega(\mathrm{P})$ must be contained in $\alpha(\gamma)$.
Theorem~\ref{futurethm} implies $\omega(\mathrm{P}) = \text{F}^0$,
hence $\text{F}^0 \in \alpha(\gamma)$. Since the saddle $\text{F}^0$
is in $\alpha(\gamma)$, the unique orbit $\phi$ emanating from it is
contained in $\alpha(\gamma)$ as well. Thus, ultimately,
$\omega(\phi)$, i.e., a point on $\text{FS}^1$, must be contained in
$\alpha(\gamma)$; this is a contradiction, since $\text{FS}^1$
consists of transversally hyperbolic sinks. We conclude that
$\gamma$ cannot contain any $\alpha$-limit point in the interior of
$\cZ^0$.

Since $\alpha(\gamma)$ must be located on the boundary on $\cZ^0$,
i.e., on $\cS_i^0$ or $\cK^0$, the proof can be completed in close
analogy to the proof in the massless case. We thus restrict
ourselves here to giving some relations that establish that the
sources on $\cZ^0$ generalize to sources on $\cX$: on
$\text{KC}_i^0$ we have $z^{-1} z^\prime |_{\text{KC}_i^0} = 2 ( 1+
\Sigma_i)$, which is positive for all $\Sigma_i> -1$ and thus for
$\Sigma_i> 1$ in particular; for $\text{QL}_i^0$ we obtain $z^{-1}
z^\prime |_{\text{QL}_i^0} = 4$. We note further that $z^{-1}
z^\prime |_{\text{D}_i^0} = 3$; thus $\text{D}_i^0$ possesses a
two-dimensional unstable manifold. (Orbits in that manifold converge
to $\text{FS}^1$.) Finally, note that along the heteroclinic cycle
$\cH_1^0: \mathrm{T}_{22}^0 \rightarrow \mathrm{T}_{32}^0
\rightarrow \mathrm{T}_{33}^0 \rightarrow \mathrm{T}_{23}^0
\rightarrow \mathrm{T}_{22}^0$, we obtain that $z^{-1} z^\prime$
equals $6 s_2$, $2(1+\Sigma_3)$, $2(1+\Sigma_2)$, $6(1-s_2)$,
respectively; hence $z^{-1} z^\prime$ is non-negative along the
heteroclinic network $\cH$.

This concludes the proof of Theorem~\ref{alphathm}.

\section{Concluding remarks}
\label{conc}

In this article we have analyzed the asymptotic behaviour of
solutions of the Einstein-Vlasov equations with Bianchi type I
symmetry. To that end we have reformulated the equations as a system
of autonomous differential equations on a compact state space, which
enabled us to employ powerful techniques from dynamical systems
theory.

Based on the global dynamical systems analysis we have identified
all possible past attractors of orbits
--- both in the massless and massive case.
We have found that an open set of solutions converges to the Kasner
circle(s); in particular, for these solutions, the rescaled matter
quantity $\Omega$ satisfies $\Omega\rightarrow 0$ toward the
singularity, so that ``matter does not matter.'' However, we have
seen that there exists an interesting structure that might
complicate matters: there exists a heteroclinic network $\cH^0$ that
might be part of the past attractor set. For solutions that converge
to $\cH^0$, $\Omega$ has no limit toward the singularity, since
$\Omega \neq 0$ along parts of the network $\cH^0$, i.e., matter
does matter for such solutions. It is not clear at the present stage
whether the set of solutions converging to $\cH^0$ is empty, or, if
non-empty, of measure zero or not (the flow on the boundary subsets
gives a hint that it might be a three-parameter set, i.e., a set of
measure zero). In any case $\cH^0$ will be important for the
intermediate dynamical behaviour of some models, and thus there are
significant differences between Bianchi type I perfect fluid models
and models with Vlasov matter.

If a generic set of orbits converges to $\cH^0$, this will have
considerable consequences. Bianchi type I perfect fluid models play
a central role in understanding the singularity of more general
spatially homogeneous models~\cite{waiell97}, as well as general
inhomogeneous models~\cite{uggetal03}. The importance of Bianchi
type I perfect fluid models is due to Lie and source
``contractions'' in spatially homogeneous cosmology and the
associated invariant subset and monotone function structure (which
is quite similar to the hierarchical structure we have encountered
in the present paper), and asymptotic silence in general
inhomogeneous models~\cite{uggetal03}. Similarly, we expect that
Bianchi type I Einstein-Vlasov models hold an equally prominent
place as regards singularities in more general
--- spatially homogeneous and inhomogeneous --- Einstein-Vlasov
models. Hence the resolution of the problem of whether the
heteroclinic network attracts generic solutions determines if
Einstein-Vlasov models are generically different from general
relativistic perfect fluid models in the vicinity of a generic
singularity, or not.

In this article we have not considered a cosmological constant,
$\Lambda$. The effects of a positive cosmological constant can be
outlined as follows: since $\rho \rightarrow \infty$ toward the
singularity, it follows that $\Lambda$ can be asymptotically
neglected and hence that the singularity structure is qualitatively
the same as for $\Lambda =0$. However, toward the future $\Lambda$
destabilizes FS$^1$, which becomes a saddle, and instead solutions
isotropize and asymptotically reach a de Sitter state.

We conclude with some remarks on different formulations. We have
seen that the variables we have used to reformulate the equations as
a dynamical system yielded multiple representations of some
structures, e.g., the Kasner circle. Replacing $s_i$ with
$E_i=\sqrt{g^{ii}}/H$, i.e., the Hubble-normalized spatial frame
variables of~\cite{uggetal03,rohugg05}, and using $y=m^2H^{-2}$
instead of $z$, yields a single Kasner circle. The latter variables,
however, are not bounded; indeed, they blow up toward the future in
the present case. It is possible to replace the variables by bounded
variables, however, variables of this type lead to differentiability
difficulties toward the singularity. Issues like these made the
variables we employed in this article more suitable for the kind of
analysis we have performed. However, $E_i$-variables, or
``$E_i$-based'' variables would have been more suitable to relate
the present results to a larger context; but it is not difficult to
translate our results to the $E_i$-variables variables used
in~\cite{uggetal03,rohugg05}, where the relationship between the
dynamics of inhomogeneous and spatially homogeneous models was
investigated and exploited.

\

\noindent
{\bf Acknowledgement}

\noindent We gratefully acknowledge the hospitality and the support
of the Isaac Newton Institute for Mathematical Sciences in
Cambridge, England, where part of this work was done. We also thank
Alan Rendall for useful questions and comments. CU is supported by
the Swedish Research Council.

\

\begin{appendix}

\section{Dynamical systems}
\label{dynsys}

In this appendix we briefly recall some concepts from the theory of
dynamical systems which we use in the article.

Consider a dynamical system defined on an invariant set $X\subseteq
\mathbb{R}^m$. The $\omega$-limit set $\omega(x)$ [$\alpha$-limit
set $\alpha(x)$] of a point $x\in X$ is defined as the set of all
accumulation points of the future [past] orbit of $x$. The simplest
examples are fixed points and periodic orbits.

The monotonicity principle~\cite{waiell97} gives
information about the global asymptotic behaviour of the dynamical
system. If $M: X\rightarrow \mathbb{R}$ is a ${\mathcal C}^1$
function which is strictly decreasing along orbits in $X$, then
\begin{subequations}\label{omegalimitmon}
\begin{align}
\omega(x) &\subseteq
\{\xi \in \bar{X}\backslash X\:|\: \lim\limits_{\zeta\rightarrow \xi} M(\zeta) \neq
\sup\limits_{X} M\} \\
\alpha(x) &\subseteq
\{\xi \in \bar{X}\backslash X\:|\:\lim\limits_{\zeta\rightarrow \xi} M(\zeta) \neq
\inf\limits_{X} M\}
\end{align}
\end{subequations}
for all $x\in X$.

Locally in the neighbourhood of a fixed point, the flow of the
dynamical system is determined by the stability features of the
fixed point. If the fixed point is hyperbolic, i.e., if the
linearization of the system at the fixed point is a matrix
possessing eigenvalues with non-vanishing real parts, then the
Hartman-Grobman theorem applies: in a neighbourhood of a hyperbolic
fixed point the full nonlinear dynamical system and the linearized
system are topologically equivalent. Non-hyperbolic fixed points are
treated in centre manifold theory: the reduction theorem generalizes
the Hartman-Grobman theorem; for further details see,
e.g.,~\cite{cra91}. If a fixed point is an element of a connected
fixed point set (line, surface,\nolinebreak \ldots) and the number
of eigenvalues with zero real parts is equal to the dimension of the
fixed point set, then the fixed point is called transversally
hyperbolic. Application of the centre manifold reduction theorem is
particularly simple in this case. (The situation is analogous in the
more general case when the fixed point is an element of an a priori
known invariant set that coincides with the centre manifold of the
fixed point.)

\section{FRW and LRS$_i$ symmetry}
\label{FRWLRS}

In this section we discuss in detail the sets $\text{FRW}$ and $\text{LRS}_i$,
connected with solutions exhibiting FRW or LRS geometry.

\begin{figure}[Ht]
\begin{center}
\psfrag{s1}[cc][cc][0.8][0]{$s_1$}
\psfrag{s2}[cc][cc][0.8][0]{$s_2$}
\psfrag{s3}[cc][cc][0.8][0]{$s_3$}
\psfrag{a}[cc][cc][0.7][0]{$v>0$}
\psfrag{b}[cc][cc][0.7][0]{$\begin{array}{cc} u<0 \\ v<0 \end{array}$}
\psfrag{c}[cc][cc][0.7][0]{$u>0$}
\psfrag{er}[cc][cc][0.7][0]{$\begin{array}{cc} u<0 \\ v=0 \end{array}$}
\psfrag{el}[cc][cc][0.7][0]{$\begin{array}{cc} u>0 \\ v>0 \end{array}$}
\psfrag{eo}[cc][cc][0.7][0]{$\begin{array}{cc} u=0 \\ v<0 \end{array}$}
\includegraphics[width=0.5\textwidth]{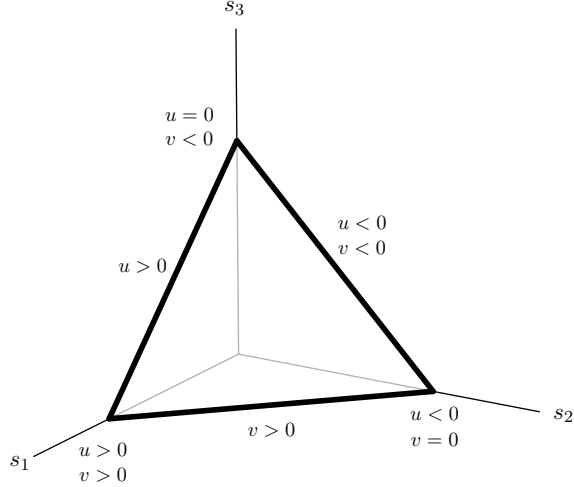}
\caption{The functions $u$ and $v$ are monotonic along the boundaries of the space
$\{(s_1,s_2,s_3)\:|\: s_i \geq 0, \sum_k s_k =1\}$.}
\label{dreiecken}
\end{center}
\end{figure}

To begin with, we prove that the fixed point $\text{F}^0$ on $z=0$
is well-defined. Since the defining equations for $\text{F}^0$ are
$w_1 = w_2= w_3 = 1/3$, we must show that these equations indeed
possess a unique solution $(s_1,s_2,s_3)$ for all distribution
functions $f_0$. Setting $z=0$ in~\eqref{omegai} implies that 
equations $w_1 = w_2= w_3 = 1/3$ are equivalent to the system
\begin{equation}\label{udef}
u := \int f_0 \, \left[s_1 v_1^2 -s_2 v_2^2\right] \left(\sum\nolimits_k s_k v_k^2 \right)^{-1/2} d^3 v =0
\end{equation}
and $v = 0$, where $v$ is defined by replacing $[s_1 v_1^2 -s_2
v_2^2]$ by $[s_1 v_1^2 -s_3 v_3^2]$ in~\eqref{udef}. On the three
boundaries of the space $\{(s_1,s_2,s_3)\:|\: s_i \geq 0, \sum_k s_k
=1\}$ the functions $u$ and $v$ are monotonic; their signs are given
in Figure~\ref{dreiecken}. The derivative $\partial u/\partial s_1$ is
manifestly positive, $\partial u/\partial s_2$ is negative, hence
$\mathrm{grad}\, u$ is linearly independent of the surface normal
$(1,1,1)$, and it follows that
$u=\mathrm{const}$ describes a curve 
for all $\mathrm{const} \in \mathbb{R}$. The same argument applies
to $v$, since $\partial v/\partial s_1>0$ and $\partial v/\partial
s_3 <0$. Figure~\ref{dreiecken} reveals that $u=0$ ($v=0$) connects
the upper (right) vertex of the $(s_1,s_2,s_3)$-space with the
opposite side. Investigating $(\mathrm{grad} \,u - \lambda\,
\mathrm{grad} \,v)$ we find that the first component is manifestly
positive when $\lambda \leq 2/3$ and negative when $\lambda\geq
3/2$, the second component is negative when $\lambda \leq 3$, and
the third component is positive when $\lambda \geq 1/3$, which
implies that $(\mathrm{grad}\, u - \lambda\, \mathrm{grad}\, v)$ is
linearly independent of the surface normal $(1,1,1)$ for all
$\lambda$. It follows that all equipotential curves of the functions
$u$ and $v$ intersect transversally; hence $u=0$ and $v=0$ possess a
unique point of intersection, which proves the claim.

By establishing existence and uniqueness of the fixed point $\text{F}^0$ for all $f_0$,
we have shown that for all distribution functions
there exists a unique FRW solution of the massless Einstein-Vlasov
equations.

The situation is different in the massive case.
A FRW solution is characterized by the equations
$\Sigma_i=0$ $\forall i$, $w_1 =w_2=w_3 =w$, since
this yields $s_{i}=\mathrm{const}$
(and a rescaling of the spatial coordinates then results in $g_{ij}\propto\delta_{ij}$.)
However, for a general distribution function $f_0$, these equations
are incompatible with 
the Einstein-Vlasov equations; in other words, the straight line
$\Sigma_i=0$ $\forall i$, $w_1 =w_2=w_3 =w$ is not an orbit of the
dynamical system. Hence, in the massive case, the Einstein-Vlasov
equations do not admit a FRW solution for arbitrary $f_0$; the
distribution function $f_0$ is required to satisfy FRW compatibility
conditions, see below, in order for a FRW solution to exist.

Note, however, that for each $f_0$, there exists exactly one orbit
that originates from $\text{F}^0$ and ends on $\text{FS}^1$, see
Section~\ref{locglo}, i.e., there exists a unique solution of
the Einstein-Vlasov equations that isotropizes toward the past and
toward the future. This anisotropic solution can be regarded as a
generalized FRW solution; if $f_0$ is compatible with the FRW
geometry, then the generalized FRW solution reduces to an ordinary
FRW solution.

The treatment of the LRS case is analogous: the subset
$\text{LRS}_1$ (and, analogously, $\text{LRS}_{2,3}$), defined
through the equations $\Sigma_2=\Sigma_3$, $w_2=w_3$, describes
solutions exhibiting LRS geometry. (For a solution on
$\text{LRS}_1$, Equation~(\ref{seq}) entails $s_2(\tau) \propto
s_3(\tau)$; by rescaling the coordinates one can achieve
$g_{22}=g_{33}$, i.e., a line element in an explicit LRS form.)
However, for general $f_0$, the set $\text{LRS}_1$ is not invariant
under the flow of the dynamical system. Consequently, for general
$f_0$, the Einstein-Vlasov equations do not admit solutions with LRS
geometry.

More specifically, consider
\begin{equation}
\left(\Sigma_2 - \Sigma_3\right)^\prime =
- 3 \Omega \left[ \textfrac{1}{2} (1-w) (\Sigma_2 -\Sigma_3) - (w_2-w_3)\right]\:.
\end{equation}
Hence, $(\Sigma_2-\Sigma_3)^\prime$ vanishes when $\Sigma_2 = \Sigma_3$ and $w_2 = w_3$.
From~\eqref{seq} and~\eqref{zeq} we obtain an equation for $w_i^\prime$,
\begin{equation}\label{omegaieq}
w_i^\prime = -2 w_i \left[ \Sigma_i -
\sum_k \Sigma_k \left(\frac{1}{2}w_k + \frac{1}{2} w_i^{-1} \beta_{i k}^{\text{\tiny (0)}}\right)
+\frac{z}{2} \left(w_i^{-1} \beta_{i}^{\text{\tiny (1)}}  + \beta^{\text{\tiny (1)}} \right) \right]\:,
\end{equation}
where we have defined
\begin{equation}
\beta_{i_1\ldots i_k}^{\text{\tiny (m)}} =
\frac{(1-z)^k {\displaystyle\int} f_0\,
\left(\Pi_{n=1}^{k} s_{i_n} v_{i_n}^2\right) \left[z+(1-z) \sum_k s_k v_k^2\right]^{1/2-k-m} d v_1 d v_2 d v_3}%
{{\displaystyle\int} f_0 \left[z+(1-z) \sum_k s_k v_k^2\right]^{1/2} d v_1 d v_2 d v_3}\:;
\end{equation}
note that $w_i = \beta_{i}^{\text{\tiny (0)}}$. Equation~\eqref{omegaieq} implies
\begin{equation}
(w_2-w_3)^\prime = -\left(\Sigma_1 -\Sigma_2\right)
\left(\beta_{22}^{\text{\tiny (0)}} - \beta_{33}^{\text{\tiny (0)}}\right) -
z \left(\Sigma_1 +1\right) \left(\beta_{2}^{\text{\tiny (1)}}- \beta_{3}^{\text{\tiny (1)}}\right)\:,
\end{equation}
when $\Sigma_2 = \Sigma_3$ and $w_2 = w_3$. We conclude that the set
$\Sigma_2 \equiv \Sigma_3$, $w_2 \equiv w_3$ is an invariant set of
the dynamical system, iff $w_2 = w_3$ implies
$\beta_{22}^{\text{\tiny (0)}} = \beta_{33}^{\text{\tiny (0)}}$ and
$\beta_{2}^{\text{\tiny (1)}} = \beta_{3}^{\text{\tiny (1)}}$. (In
the massless case, only the first condition is required.) These
conditions are violated for general distribution functions; for the
condition to hold $f_0$ must be of a certain type that ensures
compatibility with the LRS symmetry. This is the case, for instance,
when there exist constants $a_2>0$, $a_3>0$, such that $f_0$ is
invariant under the transformation $v_2 \rightarrow (a_3/a_2)\, v_3$,
$v_3\rightarrow (a_2/a_3)\, v_2$; e.g., $f_0 = \tilde{f}_0(v_1,
v_2^2 v_3^2)$, or $f_0 = \tilde{f}_0(v_1, a_2^2 v_2^2 + a_3^2
v_3^2)$; in the latter case $w_2(\tau) \equiv w_3(\tau)$ implies
$a_3^2 s_2(\tau) \equiv a_2^2 s_3(\tau)$.

Note finally that a distribution function $f_0$ is compatible with a
FRW geometry, if it is compatible with all LRS symmetries. This
means, that for instance $f_0 = \tilde{f}_0(a_1^2 v_1^2 +a_2^2 v_2^2
+a_3^2 v_3^2)$ is compatible with the FRW symmetry and thus admits a
unique FRW solution of the Einstein-Vlasov equations.

\section{Future asymptotics}
\label{futureproof}

In this section we give the proof of Theorem~\ref{futurethm}:

\noindent
\textbf{Theorem~\ref{futurethm}.}
\textit{The $\omega$-limit of every orbit in the interior of the
massive state space $\mathcal{X}$ [massless state space $\mathcal{Z}^0$]
is one the fixed points $\,\mathrm{FS}^1$ [the fixed point $\,\mathrm{F}^0$].}

\proof Consider first the state space $\mathcal{Z}^0$ of massless
particles and the associated system~\eqref{z0eq}. The function
$M_{(2)}$, cf.~\eqref{m2eqs}ff., is well-defined and monotonically
decreasing everywhere except for at the fixed point $\text{F}^0$,
where it has a global minimum. On the boundaries $\mathcal{S}_i^0$
(given by $s_i=0$) and $\mathcal{K}^0$ ($\Sigma^2 =1$) of the state
space $\mathcal{Z}^0$, the function $M_{(2)}$ is infinite.
Therefore, application of the monotonicity principle yields that the
$\omega$-limit of every orbit must be the fixed point $\text{F}^0$.

In the massive case consider~\eqref{Sigeq} in the form
\begin{equation}\label{sigmaiprime}
\Sigma_i^\prime = -3 \Omega \left[ \frac{1}{2} (1-w) (1 +\Sigma_i) - \frac{1}{2} (1- 3w) - w_i \right]\:.
\end{equation}
The r.h.s.\ is positive when $\Sigma_i \leq -1$ and $z>0$ ($w<1/3$).
This implies that the hyperplanes $\Sigma_i = -1$ constitute semipermeable membranes in the state space $\mathcal{X}$,
whereby the ``triangle'' $(\Sigma_1 > -1) \cap (\Sigma_2 > -1) \cap (\Sigma_3 > -1)$ becomes a future invariant
subset of the flow~\eqref{eq}.

The first part of the proof is to show that every orbit enters the
triangle at some time $\tau_e$ (and consequently remains inside for
all later times).

Assume that there exists an orbit with $\Sigma_i(\tau) \leq -1$ for all $\tau$ (for some $i$).
From~\eqref{seq} we infer that
\begin{equation}
s_i^\prime = -2 s_i \left[ s_j (\Sigma_i - \Sigma_j) + s_k (\Sigma_i - \Sigma_k) \right] > 0
\end{equation}
if $\Sigma_i< -1$, and that $s_i^\prime \geq 0$ if $\Sigma_i =-1$;
hence $s_i(\tau) \geq s_i(\tau_0)=\mathrm{const} > 0$ for all $\tau \in [\tau_0,\infty)$.
From~\eqref{omegaeq} we obtain
\begin{equation}
\begin{split}
\frac{1}{3} \Omega^{-1} \Omega^\prime \,\Big|_{\Omega=0} & =
1 - \frac{1}{3} w_i (1+\Sigma_i) - \frac{1}{3} w_j (1+\Sigma_j) - \frac{1}{3} w_k (1+\Sigma_k) \,\geq \\
& \geq 1 -w_j - w_k = (1-3 w) + w_i \geq \mathrm{const} >0 \:,
\end{split}
\end{equation}
since $s_i \geq \mathrm{const} > 0$. Consequently, $\Omega(\tau)
\geq \mathrm{const} > 0$ for all $\tau \in [\tau_0,\infty)$. It
follows from~\eqref{sigmaiprime} that
\begin{equation}
\Sigma_i^\prime \geq \mathrm{const} > 0
\end{equation}
for all $\tau\in [\tau_0,\infty)$ by the same argument.
This is in contradiction to the assumption $\Sigma_i \leq -1$ for all $\tau$.

Thus, in the second part of the proof, we can consider an arbitrary
orbit $\gamma$ and assume, without loss of generality, that
$\gamma(\tau)$ lies in the $\Sigma$-triangle for all $\tau \in
[\tau_e,\infty)$. Equation~\eqref{zeq} leads to
\begin{equation}
z^\prime = 2 z (1-z) \sum\limits_n s_n (1+\Sigma_n) \geq 0
\end{equation}
for all $\tau \in [\tau_e,\infty)$, hence
$z(\tau) \geq z(\tau_e) > 0$ for all $\tau \in [\tau_e,\infty)$.

We define the function $N$ by
\begin{equation}
N = (1+\Sigma_1) (1+ \Sigma_2) (1+\Sigma_3)\:.
\end{equation}
The derivative can be estimated by
\begin{equation}
N^\prime \geq 3 \Omega N \left[ -\frac{3}{2} (1-w) + \frac{1}{2} \sum_n \frac{1-3 w}{1+\Sigma_n}\right]\:.
\end{equation}
Since $w(\tau) \leq \mathrm{const} < 1/3$ (because $z(\tau) \geq
\mathrm{const} > 0$), $N^\prime$ is positive when at least one of
the $\Sigma_i$ is sufficiently small, i.e., when $N$ itself is small
(a detailed analysis shows that $N^\prime \geq 3 \Omega N [-(3/2)
(1-w) + \sqrt{3} (1-3 w) N^{-1/2}]$). We conclude that there exists
a positive constant $N_0$ such that $N(\tau) \geq N_0$ for all $\tau
\in [\tau_e,\infty)$. This in turn implies that there exists $\nu>0$
such that $\Sigma_i(\tau) \geq -1 + \nu$ for all $i$ for all $\tau
\in [\tau_e,\infty)$, whereby $z^\prime \geq 2 z (1-z) \nu$ for all
$\tau \in [\tau_e,\infty)$.

It follows that the $\omega$-limit of $\gamma$ must lie on $z=1$,
i.e., on $\mathcal{Z}^1$. Taking into account the simple structure
of the flow on $\mathcal{Z}^1$, characterized by $\Omega^\prime = 3
(1-\Omega) \Omega$, we conclude that the fixed points $\text{FS}^1$
given by $\Sigma_1 =\Sigma_2 =\Sigma_3 = 0$ are the only possible
$\omega$-limits. \proofend

\begin{remark}
In order to demonstrate the versatility of the dynamical systems
methods, we have chosen here to prove Theorem~\ref{futurethm} by
using techniques that are slightly different from those employed in
Section~\ref{locglo} (which exploit the monotonicity principle).
However, it is straightforward (in fact, even simpler) to give a
proof by making use of the hierarchy of monotone functions. Indeed,
the function $M_{(1)}$ ensures that the $\omega$-limit of every
orbit lies on $\mathcal{Z}^1$ or $\mathcal{S}_i$; modulo some
subtleties, we can exclude that $\mathcal{S}_i$ is attractive by
using the monotone function $M_{(3)}$ and the local properties of
the fixed points.
\end{remark}

\section{The spaces $\cS^0_i$ -- interpretation of solutions}
\label{Si0space}

The flow on the boundary subsets $\cS^0_i$
is of fundamental importance in the analysis of the global dynamics of the state space,
see Section~\ref{globaldynamics}. Note that except for $\text{F}^0$ all attractors
($\mathrm{D}_i^0$, $\mathrm{QL}_i^0$, $\mathrm{KC}_i^0$, and the heteroclinic network)
lie on $\cS^0_i$. For a depiction of the flow on $\cS^0_1$ see Figure~\ref{cylinder}.
In the following we show that orbits on $\cS^0_1$ represent solutions
of the Einstein-Vlasov system that are associated with a special class of
distribution functions. Furthermore, we investigate in detail solutions that
converge to the subcycle $\cH^0_1$ of the heteroclinic network.

Consider a distribution function $f_0$ of the form
\begin{equation}\label{disdisfct}
f_0(v_1,v_2,v_3) = \delta(v_1) f_0^{\mathrm{red}}(v_2,v_3)\:,
\end{equation}
where $f_0^{\mathrm{red}}(v_2,v_3)$ is even in $v_2$ and $v_3$.
In the case of massless particles, $m=0$ (and $z=0$ respectively),
we obtain
\begin{equation}
w_1 = 0 \:,\qquad w_j = \frac{g^{jj} {\displaystyle\int} f_0^{\mathrm{red}} \, v_j^2
\left[g^{22} v_2^2 + g^{33} v_3^2 \right]^{-1/2} d v_1 d v_2 d v_3}%
{{\displaystyle\int} f_0^{\mathrm{red}}
\left[g^{22} v_2^2 + g^{33} v_3^2 \right]^{1/2} d v_1 d v_2 d v_3}\:\,(j=2,3)\:,
\end{equation}
where $g^{22}$ and $g^{33}$ can be replaced by $s_2$ and $s_3$, if desired.
In the unbounded variables $g^{ii}$ the equations read
\begin{subequations}\label{unboundedvarsz0sys}
\begin{align}
\Sigma_1^ \prime & = - \Omega [1+\Sigma_1] \:,  & (g^{11})^\prime & = -2 g^{11} (1+\Sigma_1)  \\
\Sigma_j^\prime &= - \Omega [1+\Sigma_j -3 w_j] \:, & (g^{jj})^\prime &= -2 g^{jj} (1+\Sigma_j) \quad\qquad(j=2,3)\:,
\end{align}
\end{subequations}
cf.~the remark at the end of Section~\ref{einsteinvlasov}.
In particular we note that the equation for $g^{11}$ decouples; hence the full dynamics is
represented by a reduced system in the variables $(\Sigma_1, \Sigma_2, \Sigma_3, g^{22}, g^{33})$,
which coincides with the system~\eqref{unboundedvarsz0sys} on the invariant subset $g^{11} = 0$.
In analogy to the definitions~\eqref{defdimless} we set
\begin{equation}
s_1 =0 \:, \qquad s_2 = \frac{g^{22}}{g^{22}+g^{33}} \:,\qquad s_3 = \frac{g^{33}}{g^{22}+g^{33}}\:,
\end{equation}
so that $s_2 + s_3 =1$. This results in the dynamical system
\begin{subequations}\label{boundedvarsz0sys}
\begin{align}
\Sigma_1^ \prime & = - \Omega [1+\Sigma_1] \:,  & s_1 &  \equiv 0 \\
\Sigma_j^\prime &= - \Omega [1+\Sigma_j -3 w_j] \:,
& s_j^\prime &= -2 s_j [\Sigma_j - (s_2 \Sigma_2 + s_3 \Sigma_3)] \quad\qquad(j=2,3)\:.
\end{align}
\end{subequations}
This system~\eqref{boundedvarsz0sys} coincides with the dynamical system~\eqref{eq} induced on $\cS^0_1$
(which is obtained by setting $z=0$, thus $w=1/3$, and $s_1 = 0$ in~\eqref{eq}).

Our considerations show that the flow on $\cS^0_1$ possesses a direct physical interpretation:
orbits on $\cS^0_1$ represent solutions of the massless Einstein-Vlasov system
of Bianchi type~I with a ``distributional'' distribution function of the
type~\eqref{disdisfct}.
Note that the system~\eqref{boundedvarsz0sys} on $\cS^0_1$ must be supplemented
by the decoupled equations~\eqref{xeq} and $(g^{11})^\prime  = -2 g^{11} (1+\Sigma_1)$
in order to construct the actual solution from an orbit in $\cS^0_1$.

Two structures in $\cS^0_1$ are of particular interest: the fixed point $\mathrm{D}_1^0$ and
the heteroclinic cycle $\cH^0_1$, see Figure~\ref{cylinder}.
The fixed point $\mathrm{D}_1^0$ represents an LRS solution (associated with a distributional
$f_0$); it is straightforward to show that the metric is of the form
\begin{equation}\label{Disol}
g_{11} = \mathrm{const} \:,\qquad
g_{22} \propto t^{4/3} \:,\qquad
g_{33} \propto t^{4/3}\:,
\end{equation}
and $H = (4/9) t^{-1}$.

The orbit $\mathrm{T}^0_{22} \rightarrow \mathrm{T}^0_{32}$, which is part of $\cH^0_1$, corresponds to a solution
\begin{equation}\label{teil1}
g_{11} = g_{11}^0 \:, \qquad
g_{22} = g_{22}^0 ( 3 H_0 t)^2 \:, \qquad
g_{33} = g_{33}^0\:;
\end{equation}
here, $H = (3 t)^{-1}$; $H_0$ is a characteristic value of $H$.
For the orbit $\mathrm{T}^0_{33} \rightarrow \mathrm{T}^0_{23}$
the result is analogous with $g_{22}$ and $g_{33}$ interchanged.
A more extensive computation shows that the orbit $\mathrm{T}^0_{32} \rightarrow \mathrm{T}^0_{33}$
leads to
\begin{equation}\label{teil2}
g_{11} = g_{11}^0 \:, \qquad
g_{22} = g_{22}^0 \left[ \log(1+3 H_0 t) \right]^2 \:,\qquad
g_{33} = g_{33}^0 (1 +3 H_0 t)^2\:,
\end{equation}
together with $H= H_0 (1 +3 H_0 t)^{-1} (1 + [\log (1+3 H_0 t)]^{-1})$.
(Note that $3 H t$ is always close to $1$ and approaches $1$ for $t \rightarrow 0$
and $t\rightarrow \infty$.)
The result for the orbit $\mathrm{T}^0_{23} \rightarrow \mathrm{T}^0_{22}$ is analogous with
$g_{22}$ and $g_{33}$ interchanged.

Now consider an orbit converging to the heteroclinic cycle as $\tau \rightarrow -\infty$, i.e., $t \searrow 0$.
Since the orbit alternates between episodes where it is close to one of the four heteroclinic orbits,
we obtain a solution with alternating episodes of characteristic behaviour of
the type~\eqref{teil1} and~\eqref{teil2}; transitions between the episodes
correspond to the orbit being close to the fixed points.

Let $t^{(n)}$ denote a monotone sequence of times such that
the solution is in episode $(n)$ at time $t^{(n)}$
(i.e., the orbit is close to one of the four heteroclinic orbits and
far from the fixed points); $t^{(n)}\searrow 0$ as $n\rightarrow \infty$.
Since $3 H t \approx 1$ as $t\searrow 0$, the sequence $t^{(n)}$ gives rise
to a sequence $H^{(n)}$ defined by $3 H^{(n)} t^{(n)} = 1$.
During episode $(n)$ the solution exhibits a characteristic behaviour of
the type~\eqref{teil1} or~\eqref{teil2} with $H_0 =H^{(n)}$
(and $g_{22}^0 = g_{22}^{(n)}$, $g_{33}^0 = g_{33}^{(n)}$).
A transition from one episode to another involves
a matching of the constants.

\textit{Example}.
Suppose that the orbit is close to the heteroclinic orbit
$\mathrm{T}^0_{32} \rightarrow \mathrm{T}^0_{33}$ in episode $(n)$.
We obtain a behaviour of the type~\eqref{teil2} with $H_0=H^{(n)}$.
As $H^{(n)} t$ gets small we see that $g_{22} \approx g_{22}^{(n)} (3 H^{(n)} t)^2$,
$g_{33} \approx g_{33}^{(n)}$.
The next (as $t\searrow 0$) episode corresponds to the orbit running close
to $\mathrm{T}^0_{22} \rightarrow \mathrm{T}^0_{32}$;
the behaviour of the solution is~\eqref{teil1} with $g_{22}^{(n+1)}$, $g_{33}^{(n+1)}$,
and $H_0=H^{(n+1)}$.
The transition between the episodes $(n)$ and $(n+1)$ is thus straightforward:
$g_{22}^{(n+1)} (H^{(n+1)})^2 = g_{22}^{(n)}  (H^{(n)})^2$ and
$g_{33}^{(n+1)} = g_{33}^{(n)}$.
Matching episodes $(n+1)$ and $(n+2)$ is slightly more involved.
The orbit is close to the heteroclinic orbit $\mathrm{T}^0_{23} \rightarrow \mathrm{T}^0_{22}$
in episode $(n+2)$, where
\begin{equation}\label{teil4}
g_{11} = g_{11}^0 \:, \qquad
g_{22} = g_{22}^{(n+2)} (1 +3 H^{(n+2)} t)^2 \:,\qquad
g_{33} = g_{33}^{(n+2)} \left[ \log(1+3 H^{(n+2)} t) \right]^2 \:.
\end{equation}
Close to transition time, when $H^{(n+2)} t$ is large,
we get $g_{22} = g_{22}^{(n+2)} (3 H^{(n+2)} t)^2$ and $g_{33} = g_{33}^{(n+2)} (\log 3 H^{(n+2)} t)^2$.
The transition between episode $(n+1)$ and $(n+2)$ thus involves
that $g_{33}$ begins to decay logarithmically from having been approximately
constant.

\end{appendix}

\end{document}